\documentclass[apj]{emulateapj}
\usepackage{epsf, graphicx, subfigure}
\usepackage{amsmath} 
\usepackage{natbib}
\usepackage{color}

\usepackage{colortbl}
\usepackage{hyperref}
\hypersetup{
    colorlinks,%
    citecolor=blue,%
    filecolor=black,%
    linkcolor=blue,%
    urlcolor=black
}

\renewcommand{\arraystretch}{1.2}

\shorttitle{Star-disk Collisions in the GC}
\shortauthors{Kieffer \& Bogdanovi\'c}

\begin{document}

\title{Can Star-Disk Collisions Explain the Missing Red Giants Problem in the Galactic Center?}
\author{T. Forrest Kieffer\altaffilmark{1}\footnote{Barry Goldwater Scholar} \& Tamara Bogdanovi\'c\altaffilmark{2}}
\affil{Center for Relativistic Astrophysics, School of Physics \\ Georgia Institute of Technology, Atlanta, Georgia 30332}
\email{tkieffer3@gatech.edu, tamarab@gatech.edu}

\begin{abstract}
Observations have revealed a relative paucity of red giant (RG) stars within the central 0.5\,pc in the Galactic Center (GC). Motivated by this finding we investigate the hypothesis that collisions of stars with a fragmenting accretion disk are responsible for the observed dearth of evolved stars. We use 3-dimensional hydrodynamic simulations to model a star with radius $10 R_{\odot}$ and mass $1 M_{\odot}$, representative of the missing population of RGs, colliding with high density clumps. We find that multiple collisions with clumps of column density $\gtrsim10^{8}\, {\rm g\,cm^{-2}}$ can strip a substantial fraction of the star's envelope and in principle render it invisible to observations. Simulations confirm that repeated impacts are particularly efficient in driving mass loss as partially stripped RGs expand and have increased cross sections for subsequent collisions. Because the envelope is unbound on account of the kinetic energy of the star, any significant amount of stripping of the RG population in the GC should be mirrored by a systematic decay of their orbits and possibly by their enhanced rotational velocity. To be viable, this scenario requires that the total mass of the fragmenting disk has been several orders of magnitude higher than that of the early type stars which now form the stellar disk in the GC.
\end{abstract}

\section{Introduction}
\label{sec:intro}

Encounters of stars with a nuclear accretion disk are likely to play an important role in the appearance and evolution of nuclear regions in galaxies. In this process, both the accretion disk and the nuclear star cluster (NSC) may be affected. For example, gravitational torquing and impacts by the cluster stars can enhance angular momentum transport in the nuclear accretion disks \citep{Ost83},  produce bright hot spots \citep{Zen83} and result in hotter disks \citep{NorSil83,PerWil93}. Stellar impacts can also lift off filaments of gas from the disk surface which have been proposed as the origin of the broad emission lines in quasars \citep{ZurSie94}.

On the other hand, cluster stars can lose orbital energy and angular momentum in each collision with the disk causing the systematic decay of their orbits around the central supermassive black hole \citep[SMBH;][]{Rau95, KarSub01}. It was found for example that the star-disk interactions tend to order stellar orbits, dragging them into the disk plane, while star-star collisions tend to scatter them \citep{vilkoviskij2002role}. Stars that loose a sufficient amount of orbital energy and momentum can be brought to co-rotation with the disk and continue to spiral in until they are tidally disrupted or collide with other stars \citep{SyeCla91}. If a star in question happens to be a white dwarf, it could gain a substantial amount of mass through accretion and burst repeatedly as a nova \citep{shields1996extraordinary}. Alternatively, compact objects may spiral in to the center of the cluster and coalesce with the SMBH through emission of gravitational waves. 

Given a rich phenomenology of star-disk interactions it is interesting to consider whether any of these phenomena operate in our own Galactic Center (GC), which is one of the best laboratories for high resolution studies of nuclear stellar dynamics. Stellar impacts for example have been proposed as an explanation for the luminous X-ray flares \citep{NayCua04, DaiFue10} that have been observed in the GC \citep{BagBau01}. It has also been suggested that star-disk interactions can have important implications for the properties and appearance of the NSC in the GC \citep{GheSal05, GenKar07, GilEis09}.  

These studies provide a broader context for the question investigated in this paper, which pertains to the impact of star-disk interactions on the population of red giant (RG) and horizontal branch (HB) stars in the GC. The motivation for this investigation stems from observations that indicate a relative paucity of evolved stars and a high concentration of younger, hot blue stars in the central half-parsec \citep{1991ApJ...382L..19K, 1994A&A...285..573N, BuhhSch09, DoGhez09, Bar10,stosad15}. Down to their limiting magnitude ($K_s\sim 18$) these surveys indicate that the missing late type stars comprise RGs and HBs with age $\gtrsim 1$ Gyr and mass in the range $0.5-4\,M_{\odot}$ \citep{genzel10}.
For brevity, we will only refer to the RG population in the remainder of the manuscript while keeping in mind both populations of evolved stars. 

If the early and late type stellar populations evolved under similar conditions, both should exhibit a characteristic, cuspy distributions centered on the SMBH. The indication that the two distributions differ however points to a mechanism that can create a core in the observed population of RGs, by either removing them physically or by rendering them unobservable. Proposed explanations for the dearth of RGs within the central parsec of the GC include star-star collisions \citep{GenTha96, DavBla98, BaiDav99, Ale99, DalDav09}, MBH binaries scouring out a core in the GC via three-body interactions \citep{BauGua06, PorBau06, MatMak07, LocBau08, GuaMer12}, and infalling star clusters \citep{KimMor03, ErnJus09, AntCap12}. All of these mechanisms have encountered varying degrees of difficulty in explaining the properties of the observed stellar populations in the GC.

More recently, \citet{AmaChe13} (hereafter, ASC) proposed that the observed distribution of the RG stars can be a consequence of the star-disk collisions. This hypothesis is motivated by the observational evidence for a disk of early-type stars surrounding the SMBH and extending from approximately $0.04$\,pc out to $0.5$\,pc \citep{PauGen06, LuGhe09, bartko09,stosad15}. These are Wolf-Rayet and O-type (WR/O) stars with the age of only several Myr and masses in the range $10-60\,M_{\odot}$. The existence of the stellar disk is indicative of in-situ star formation which presumably started with fragmentation of a gravitationally unstable nuclear gas disk \citep{LevBel03, levin07, AleArm08}. In the star-disk collision scenario the RGs in the NSC collide with dense clumps in the fragmenting disk. Because they have compact cores surrounded by tenuous outer layers, RGs are particularly vulnerable to collisions, which can lead to large amounts of mass loss from the star. It follows that collisions of RGs with the fragmenting accretion disk could potentially render them ``invisible" to observations or even completely disrupted.

In this work we test the hypothesis that the missing RGs are a result of impacts with a fragmenting disk via high resolution hydrodynamic simulations. This investigation is preceded by that of \cite{ArmZur96} who used smoothed particle hydrodynamic simulations to study whether stripped RGs are a significant source of fuel for SMBHs in active galactic nuclei (AGN). They show that RGs with $R_* \approx 150 R_{\odot}$ can in some cases be completely stripped of their outer envelope leaving only the central high density core, in agreement with the more recent analytic results of ASC. 

Here we focus on more compact stars with radii $R_* \approx 10 R_{\odot}$ and mass $M_* \approx 1M_{\odot}$ that are representative of the missing population of RG and HB stars in the GC. Because a smaller stellar radius implies higher binding energy of the envelope, evolved stars with these properties are harder to strip and disrupt. As a consequence, significant mass loss for the RGs in the GC can occur only in the late stages of nuclear disk fragmentation and after multiple collisions with high density clumps. Given these more stringent criteria motivated by our simulations, we re-evaluate the hypothesis that star-disk collisions are responsible for the dearth of the late type stars in the central 0.5\,pc of our Galaxy.

This paper is organized as follows: in \S \ref{sec:NM} we give an overview of the numerical methods and initial conditions for the star and disk configuration. In \S \ref{sec:results} we present the results of the study, in \S \ref{sec:Discussion} discuss their implications and limitations and conclude in \S \ref{sec:conc}. 


\section{Numerical Setup}
\label{sec:NM}

We construct models in a Cartesian coordinate system $(x, y, z)$ with a cubic spatial domain centered on the star. The size of the domain is defined by $x=\pm 2R_*$, $y=\pm 2R_*$, $z=\pm 2R_*$, where $R_*$ is the radius of the star. A base numerical resolution used in the majority of the simulations is $128^3$ unless noted otherwise. All boundary surfaces use a zero-gradient outflow boundary condition (i.e., the fluid may freely flow out) except for the surface through which matter, representing a local region of the accretion disk, flows into the domain. In this setup, we are essentially placing the RG in a ``wind tunnel". Before impacting the disk the star is momentarily immersed in a low density atmosphere,  $\rho_{\text{a}} = 10^{-15} \rho_{c,0}$ where $\rho_{c,0}$ is the central density of the RG at the beginning of the simulation.

To simulate the star-disk encounters we use a version of the hydrodynamic code VH-1 \citep{VH-1} to solve the equations for inviscid flow of an ideal compressible gas. VH-1 is a grid based parallel code that combines a three-dimensional finite difference approach for hydrodynamics and a spectral co-location technique for the self-gravity \citep[see][for description of the implementation, tests and applications]{ChengEva13}. The code is based on the Piecewise Parabolic Method \citep[PPM;][]{Col84} and uses the Lagrangian-remap formulation of the method.

\begin{figure*}[t] 
\epsscale{1.0}  
\plottwo{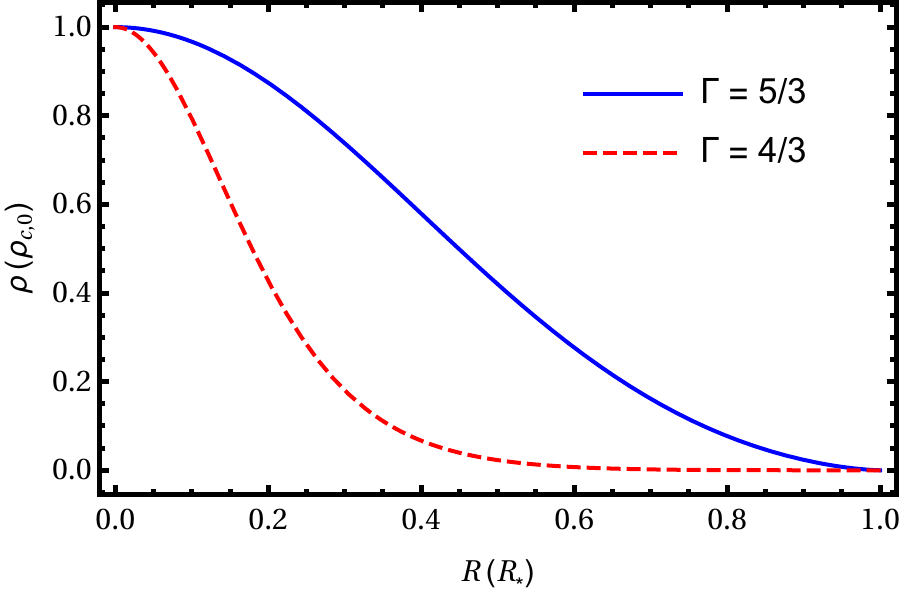}{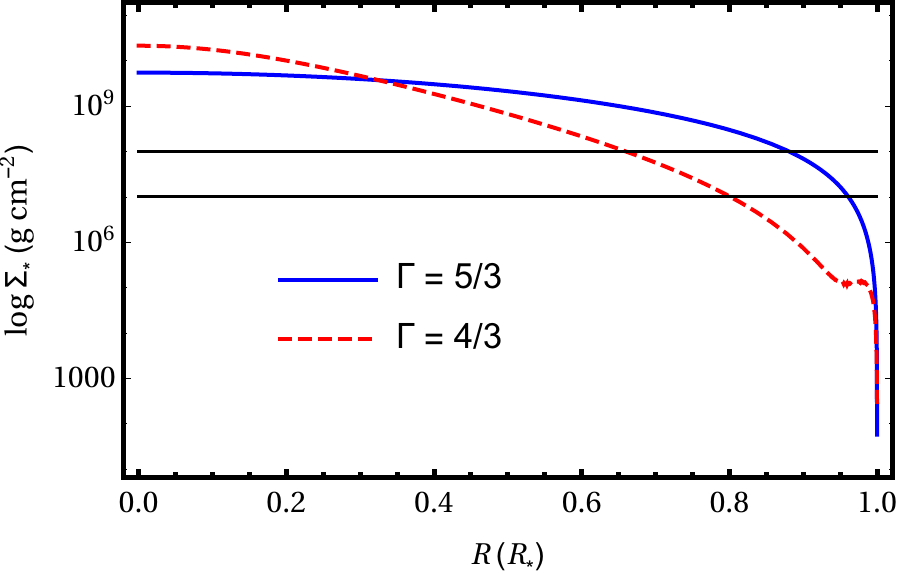}
\caption{{\it Left:} Initial density profiles of simulated RGs: $\Gamma = 5/3$ (blue, solid) and $4/3$ (red, dashed). {\it Right:} Initial column density profiles for the same two models. Horizontal lines at $10^7$ and $10^8$ g cm$^{-2}$ correspond to column densities of simulated clumps. The "kink" in the $\Gamma = 4/3$ profile close to the surface of the star is a numerical artifact (see the text).}
\label{fig:AdiabaticIndex}
\end{figure*}

\subsection{Properties of the star}
\label{sub:SettingupRG}

A modeled star has the initial radius $R_{*} = 10 R_{\odot}$ and mass $M_{*} = 1 M_{\odot}$, similar to the properties of the missing giants in the GC. We describe it as a self-gravitating atmosphere in hydrostatic equilibrium with a polytropic equation of state, $P = K \rho^{\Gamma}$, where $K$ is the polytropic constant, $\Gamma = 1 + 1/n$ is the adiabatic index, and $n$ is the polytropic index. The initial density profiles of the stars are constructed by numerically solving the Lane-Emden equation for $n = 3/2$ and $n = 3$, corresponding to $\Gamma = 5/3$ and $\Gamma = 4/3$, respectively \citep{1967aits.book.....C}. Regardless of the structure of the star we evolve the properties of the gas everywhere in the computational domain using the equation of state of ideal gas and a fixed adiabatic index $\gamma = 5/3$. 
 
The left panel of Figure~\ref{fig:AdiabaticIndex} shows the initial density profiles for $\Gamma = 5/3$ (solid) and $4/3$ (dashed) models. Both are normalized to the initial central density of the star which for $\Gamma = 5/3$ ($\Gamma = 4/3$) corresponds to $\rho_{c,0} = 8.4 \times 10^{-3}\,{\rm g\,cm^{-3}}$ ($\rho_{c,0} = 7.6 \times 10^{-2}\,{\rm g\,cm^{-3}}$).  Note that the central densities of RG stars are in reality much higher ($\sim 10^5\,{\rm g\,cm^{-3}}$) within the central $\sim 10^9 {\rm cm}$ in the stellar core. Our polytropic stellar models do not capture this peak in the central density but can rather be thought of as the density distribution of an extended envelope in hydrostatic equilibrium. We discuss further the implications of this choice in \S~\ref{sec:Discussion}.

The right panel of Figure~\ref{fig:AdiabaticIndex} shows the initial column density profiles for the same two models. The "kink" in the $\Gamma = 4/3$ column density profile that appears close to the star surface is the artifact that arises from the polynomial interpolation and integration of the numerical solution to the Lane-Emden equation. Since the mass of the stellar envelope enclosed in the kink region is negligible, we conclude that this artifact does not affect the stability of the modeled star or results of our simulations. 

While neither polytropic profile is a precise description of the RG structure, together they outline a family of models representative of the low mass stars ($M_{*} \sim 0.1 M_{\odot}$, $\Gamma = 5/3$) to moderate mass stars with compact cores ($M_{*}\sim 1 M_{\odot}$, $\Gamma = 4/3$). A majority of our simulations focus on the polytropic model $\Gamma = 5/3$ due to the numerical facility, since simulation of a $\Gamma = 4/3$ model requires numerical resolution higher than $256^3$ in order to resolve the steep pressure and density gradients in the compact core region at radii $<0.3R_*$ (left panel of Figure~\ref{fig:AdiabaticIndex}).  This makes $\Gamma = 4/3$ models $>2^4$ times more computationally expensive relative to the remainder of the simulations in this study. We discuss the convergence of our models as a function of numerical resolution in more detail in the Appendix. In Table~\ref{tab:ParameterSpace} we name $\Gamma = 5/3$ runs with the letter ``A" and $\Gamma = 4/3$ runs with ``B". The second component in the run names marks numerical resolution, when it is different from the baseline resolution of $128^3$. For example, run A7\_256 corresponds to the $\Gamma = 5/3$ model and numerical resolution of $256^3$.

We assume that the velocity of a star-disk encounter is comparable to the orbital velocities of the stars about the center of the NSC, which within the radius of $r\leq 0.5$\,pc correspond to $v_* = (GM_\bullet/r)^{1/2} \gtrsim 200\,{\rm km\,s^{-1}}$ \citep{trippe2008kinematics}. Here we account for the mass of the black hole in the GC \citep[$M_\bullet = 4.3\times 10^6\,M_\odot$;][]{ghez08,GilEis09} and neglect the contribution due to the mass of the enclosed NSC stars, which at $r\leq 0.5$\,pc is about an order of magnitude lower than that of the SMBH \citep{sme09}. We therefore simulate the initial encounter velocities in the range from 150 to  $1200\, {\rm km\, s^{-1}}$, as outlined in Table~\ref{tab:ParameterSpace}. 

Note that the orbital velocity of the RG decreases over the course of the simulation due to the exchange of linear momentum with the accretion disk. We continuously adjust the reference frame of the computational domain so that the star remains at rest and centered on the domain throughout the simulation.


\begin{deluxetable*}{cccccccccc} 
\tabletypesize{\footnotesize} 
\tablecolumns{11} 
\tablewidth{0pt} 
\tablecaption{Simulation parameters. $\Gamma$ -- polytropic index. $N_{\rm res}$ -- numerical resolution. $v_*$ -- velocity of the star. $\Sigma_c$ -- clump column density. $R_c$ -- clump radius. $\mathcal{M}$ -- Mach number. $t_{\rm cc}$ -- clump crossing time.  $t_{\rm sim}$ -- simulation length. $N_{\rm coll}$ -- number of collisions. \label{tab:ParameterSpace}} 
\tablehead{ 
Run  & $\Gamma$ & $N_{\rm res}$ & $v_{*}$ & $\Sigma_c$ & $R_c$ &  $\mathcal{M}$ & $t_{\rm cc}$ & $t_{\text{sim}}$ & $N_{\rm coll}$  \\
& & & (km\,s$^{-1})$  & (g\,cm$^{-2}$)  & (cm) & & ($t_{\text{dyn}}$)  &  ($t_{\text{dyn}}$) & }
\startdata 
A1 & 5/3 & 128 & 300 & $10^7$ & $7.9  \times 10^{13}$  & 1.8 & 52 & 100 & 2 \\ 
RA1 & 5/3 & 128 & 300 & $10^7$ &   $7.9  \times 10^{13}$ & 1.8 &  52 & 300 & 2 \\ 
B1 & 4/3 & 128 & 300  & $10^7$ &   $7.9  \times 10^{13}$ & 1.8 & 52 & 100 & 2 \\ 
B1\_256 & 4/3 & 256 & 300  & $10^7$ &   $7.9  \times 10^{13}$ & 1.8 & 52 & 100 & 2 \\ 
B1\_300 & 4/3 & 300 & 300  & $10^7$ &   $7.9  \times 10^{13}$ & 1.8 & 52 & 100 & 2 \\ 
B1\_512 & 4/3 & 512 & 300  & $10^7$ &   $7.9  \times 10^{13}$ & 1.8 & 52 & 44 & $<$1  \\  
A2 & 5/3 & 128 & 600  & $10^7$ &    $7.9  \times 10^{13}$ & 3.6 & 26 & 100  & 4 \\ 
A3 & 5/3 & 128 & 900  & $10^7$ &  $7.9  \times 10^{13}$ & 5.4 & 17 & 100 & 6  \\ 
A4 & 5/3 & 128 & 1200  & $10^7$ &  $7.9  \times 10^{13}$ & 7.1 & 13 & 100  & 7.5 \\ 
A5 & 5/3 & 128  & 150 & $10^8$ & $2.5 \times 10^{13}$  & 0.5 & 33  & 100 & 3 \\
RA5 & 5/3 & 128  & 150 & $10^8$ & $2.5 \times 10^{13}$ & 0.5 & 33 & 293  & 3  \\
A6 & 5/3 & 128 &  300 & $10^8$ & $2.5 \times 10^{13}$ & 1.0 & 16 & 100 &  6 \\
A7 & 5/3 & 128 &  600 & $10^8$ &  $2.5 \times 10^{13}$ & 2.0 & 8 & 100 & 12.5 \\
A7\_64 & 5/3 & 64 &  600 & $10^8$ &  $2.5 \times 10^{13}$ & 2.0 & 8 & 100 & 12.5  \\
A7\_256 & 5/3 & 256 &  600 & $10^8$ &  $2.5 \times 10^{13}$ & 2.0 & 8 & 100 & 12.5 \\
RA7 & 5/3 & 128 &  600 & $10^8$ &  $2.5 \times 10^{13}$ & 2.0 & 8  & 265  & 12 \\
B7\_300 & 4/3 & 300 &  600 & $10^8$ &  $2.5 \times 10^{13}$ & 2.0 & 8 & 100 & 12 \\
A8 & 5/3 & 128 &  900 & $10^8$ & $2.5 \times 10^{13}$ & 3.0 &  5 & 100  & 20  \\
A9 & 5/3 & 128 &  1200 & $10^8$ & $2.5 \times 10^{13}$ & 4.0 & 4 & 100 & 25 
\enddata 
\tablecomments{Note that $N_{\rm coll}$ is the same for corresponding continuous and repeated impact runs so that in both scenarios the star spends equal time traveling within the clump.} 
\end{deluxetable*}


\subsection{Properties of the Fragmenting Accretion Disk}
\label{sub:runs}

We assume that just before fragmentation the nuclear accretion disk in the GC is characterized by the critical value of the Toomre parameter, $Q\approx1$, and has properties as described by \citet{levin07}. Assuming that it has the same spatial extent as the young stellar disk presently observed in the GC (see Section~\ref{sec:intro}), the surface density and half-height of such fragmenting disk are
\begin{eqnarray}
\Sigma_{0.04}  \simeq  200\, {\rm g\,cm^{-2}} \hspace{2mm}  & {\rm and}\hspace{2mm} &
\Sigma_{0.5}  \simeq  1\, {\rm g\,cm^{-2}}\label{eq:diskcolumn} \\
h_{0.04}  \simeq  1.4 \times 10^{14}\, {\rm cm}  & \hspace{2mm} {\rm and}\hspace{2mm} &
  h_{0.5}  \simeq  1.3 \times 10^{15}\, {\rm cm}\label{eq:diskheight}
\end{eqnarray}
where subscripts 0.04 and 0.5 in parsecs refer to the values at the inner and outer disk edge, respectively. Because the disk is relatively cold ($T\approx10-10^2$\,K), its opacity is dominated by light scattering off ice grains and in some cases by scattering off metal dust. \citet{levin07} finds that such marginally self-gravitating disk can reach the total mass of $\sim {\rm few}\times 10^4\,M_\odot$, is capable of forming clumps with mass $\sim10^2 \,M_\odot$ and possibly up to $10^3 \,M_\odot$, if massive clumps can avoid opening a gap in the disk from which they are assembling.

Because of its relatively low surface density (compared to the star), the encounter with the disk of such properties leaves the RG unscathed.  We choose a similar setup as a numerical test of the stability of the star placed in a low density background flow (see Appendix) and in the remainder of the paper consider collisions of the RG star with the higher density clumps.

The initial properties of the accretion disk imply that clumps that form from it as a consequence of fragmentation and runaway collapse of gas clouds must have radii {\it smaller} than the disk half-thickness $h$ 
\begin{equation}
R_c \simeq \min{ \{ \sqrt{M_c / \pi \Sigma_c } , h \} }
\label{eq_Rc}
\end{equation}
where $R_c$, $M_c$ and $\Sigma_c$ are the clump radius, mass, and surface density, respectively. It also follows that  the surface density of the clump must be larger than that of the disk initially. We choose clump surface densities $\Sigma_c = 10^7$ and $10^8$ g cm$^{-2}$, for which the RG losses a non-negligible amount of its outer envelope due to stripping (Figure~\ref{fig:AdiabaticIndex}). Following the approach by ASC, who conjecture that the clumps must be sufficiently massive to produce the WR/O stars in the observed stellar disk in the GC, and consistent with the estimates by \citet{levin07}, we adopt the clump mass of $M_c = 100\, M_{\odot}$. We use equation~\ref{eq_Rc} to estimate the clump radius, $R_c$, and list the corresponding values in Table~\ref{tab:ParameterSpace}.

Assuming that collapsing clumps evolve through a sequence of hydrostatic equilibria, where at every instance the gravity and thermal pressure are in near balance, we estimate the sound speed in the clumps as $c_s \approx (\gamma k_B T_c /m_p)^{1/2} \approx (\gamma GM_c /R_c)^{1/2}$, where $T_c$ is the clump temperature and the constants have their usual meaning. With the central temperature on the order of $\sim 10^6$\,K and characteristic size of several AU, such clumps are veritable protostars at the verge of deuterium burning \citep{masunaga00,bate14}. The resulting sound speeds are about 170 and 300\,${\rm km\,s^{-1}}$ for the clumps with surface densities $10^7$ and $10^8$ g cm$^{-2}$, respectively. Given the range orbital velocities of stars in the central 0.5\,pc (see \S~\ref{sub:SettingupRG}), it follows that for this choice of parameters a majority of star-disk encounters will be supersonic, as illustrated by the value of the Mach number, $\mathcal{M} = v_*/c_s$, shown in Table~\ref{tab:ParameterSpace}.

In simulations, we model clumps as uniform density slabs and neglect their inner structure (e.g., density gradients and granularity). The clump is introduced as a continuous inflow of gas from one boundary of the computational domain with a velocity determined by the orbital velocity of an RG. In a subset of simulations, this inflowing gas is continually injected into the domain, simulating a star impacting a clump once and never exiting. However, in order to investigate the effect of repeated impacts, simulations are also carried out in which the RG travels through multiple, discrete slabs of gas. The thickness of each slab is determined by the clump crossing time (see equation \ref{eq_tc}, \S \ref{sub:time}). The examples of ``continuous" and ``repeated impact" simulations are the runs A7 and RA7, respectively.  

\subsection{Characteristic time scales}
\label{sub:time}

We use the dynamical time as a natural time unit in our simulations
\begin{equation}
t_{\text{dyn}} = \frac{R_*^{3/2}}{(G M_*)^{1/2}} \simeq
11.4\,{\rm h} \left(\frac{R_*}{10 R_\odot}\right)^{3/2}
\left(\frac{M_*}{M_\odot}\right)^{-1/2}
\end{equation}
which for a star in hydrostatic equilibrium is comparable to its sound crossing time scale. The time scale that describes thermal evolution of the star is given by its Kelvin-Helmholtz time scale. For the population of late-type giants in the GC, with temperatures in the range $\sim 3500-3700$\,K \citep{genzel10}, it amounts to
\begin{equation}
t_{\text{KH}} = \frac{G M_*^2} {R_* L_*}\simeq
3.2\times10^5\,{\rm yr} 
\left(\frac{M_*}{M_\odot}\right)^2
\left(\frac{R_*}{10 R_\odot}\right)^{-1}
\left(\frac{L_*}{10 L_\odot}\right)^{-1}
\label{eq_tkh}
\end{equation}
where $L_*$ is the RG luminosity. The orbital period of a star at the radius $r$ from the center of the cluster is 
\begin{equation}
t_{\rm orb} = \frac{2 \pi r}{v_*} \simeq 1.6\times 10^4\,{\rm yr}\;
r_{0.5}^{3/2}\, M_\bullet^{-1/2}
\label{eq_torb}
\end{equation}
where $r_{0.5}$ is in units of 0.5\,pc, and $M_\bullet$ is the mass of the black hole in the GC. For these values, $t_{\rm orb}$ in equation~\ref{eq_torb} corresponds to $\sim 10^7\, t_{\rm dyn}$. Given that $1 M_{\odot}$ star spends $t_{\rm rg} \sim 10^8$\,yr in the RG and HB phase of evolution \cite[see Figure~1 in][]{macleod12} the number of orbits that stars within the central 0.5\,pc of the NSC complete during this time can be $t_{\rm rg} / t_{\rm orb} \gtrsim 3\times 10^4$. This simple estimate does not take into account the evolution in the orbital period nor the fact that the star may be disrupted by the SMBH or as a consequence of collisions with the disk. 


\begin{figure*}[th!]
\includegraphics[scale=0.45]{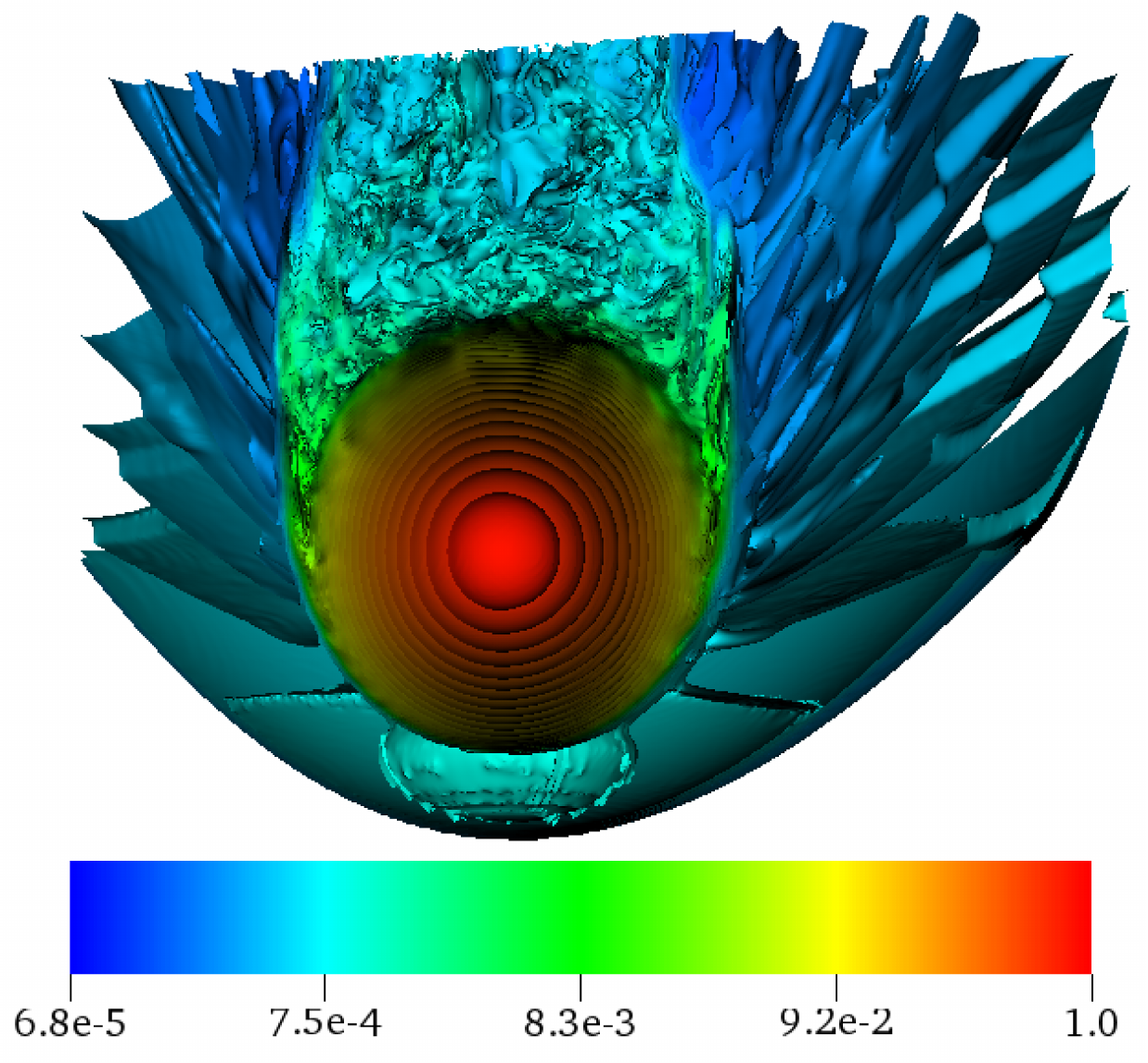} 
\includegraphics[scale=0.45]{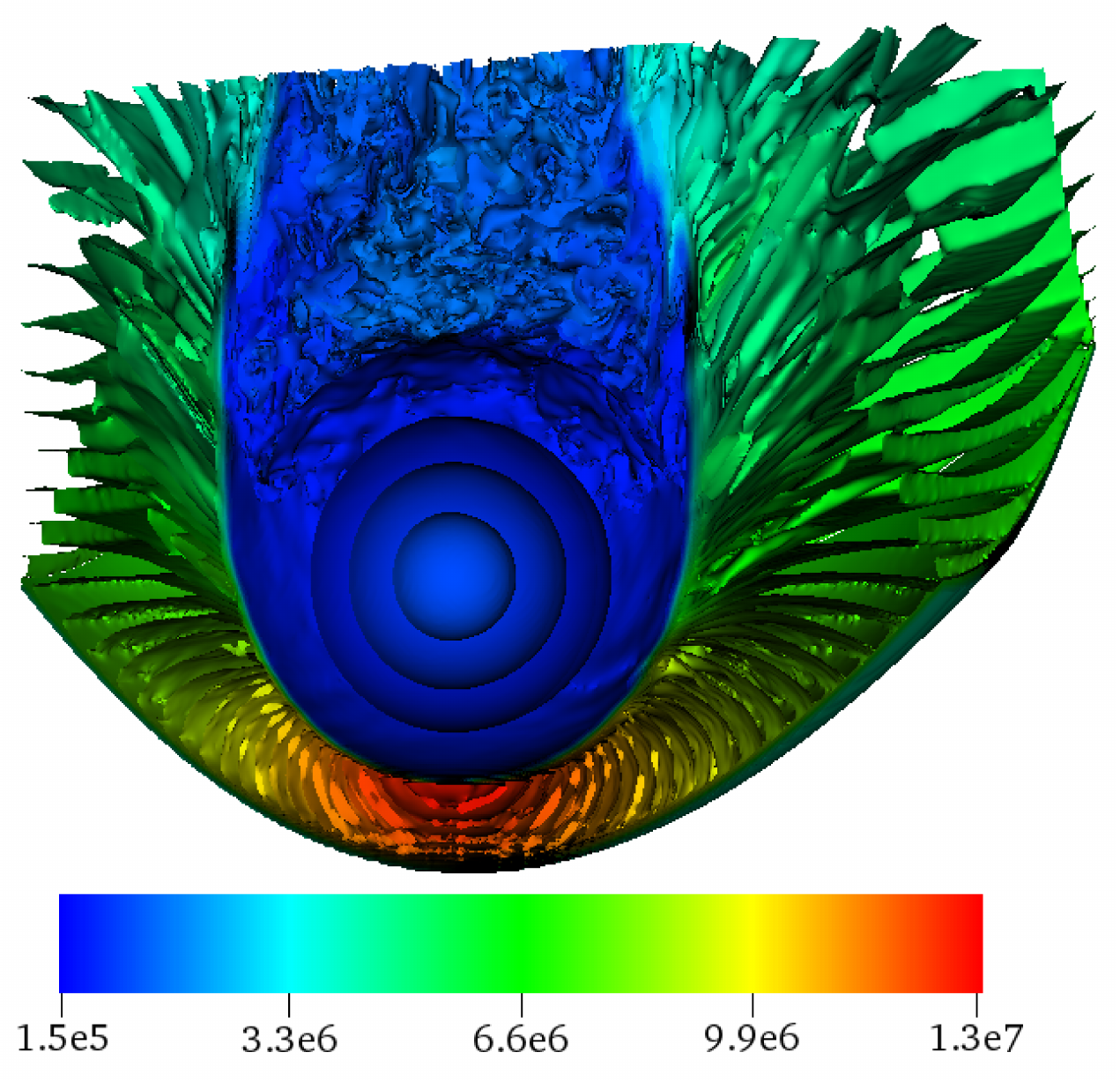}  
\includegraphics[scale=0.45]{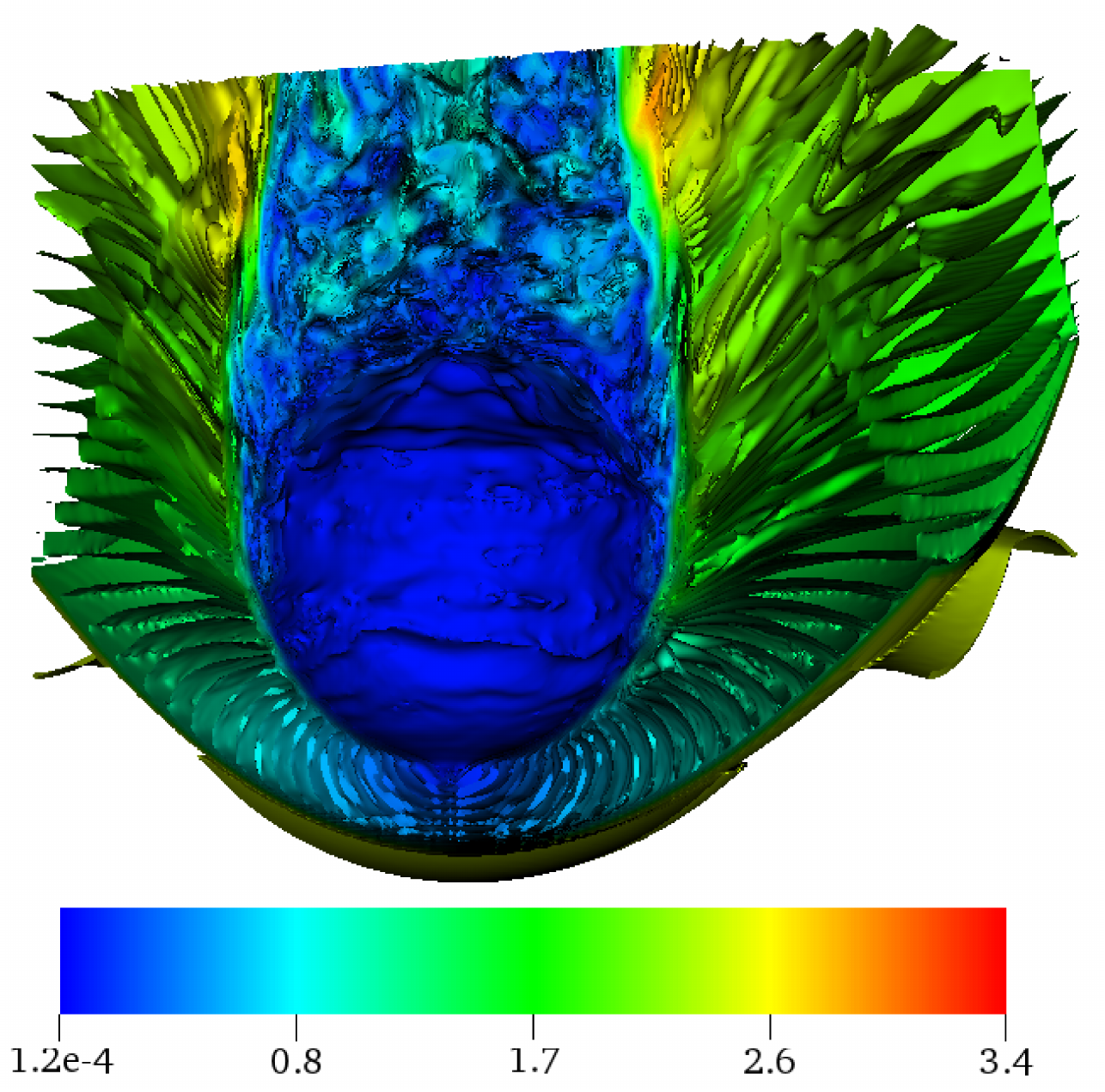}
\caption{Three-dimensional isosurfaces of density (left), temperature (middle), and the Mach number (right) from run A7\_256 taken at $10\, t_{\rm dyn}$ after the initial impact. Fluid that represents the clump is moving upwards as indicated by the shape of the bow shock. Color bars below the density and temperature panels show values in units of $\rho_{c,0}$ and Kelvin, respectively.}
\label{fig:VisIt_Temp}
\end{figure*}


Over the course of each orbit the star will only spend a fraction of time interacting with the clumps in the fragmenting disk.  We refer to this time scale as the clump crossing time and express it as 
\begin{equation}
t_{\rm cc} = \frac{R_c}{v_*} \simeq 1.7\times 10^{-2}\,{\rm yr}\;
R_{c,13}\, r_{0.5}^{1/2}\, M_\bullet^{-1/2} .
\label{eq_tc}
\end{equation}
where $R_{c,13} = R_c/10^{13}\,{\rm cm}$. The clump crossing time for each simulation is shown in Table~\ref{tab:ParameterSpace}.  For a full range of the clump properties and RG orbital velocities considered here $t_{\rm cc}$ ranges from 4 to $50\, t_{\rm dyn}$.

Another relevant time scale is the life time of a clump, during which the clump collapses to form a protostar. According to models of the protostellar collapse \citep{masunaga00, bate14} the slowest process that dominates the life time of a collapsing clump is the assembly of the initial clump core up to the point when its central temperature reaches the dissociation temperature for molecular hydrogen. This time scale is well described by the free-fall time of the initial clump core, 
\begin{equation}
t_{\rm lc} \approx \sqrt{\frac{3\pi}{32 G\rho}} \sim 
2400\,{\rm yr}\,\Sigma_{0.5}^{-1/2}\, h_{0.5}^{1/2}
\label{eq_tlc}
\end{equation}
where for the purpose of this estimate we used properties of the fragmenting disk at its outer edge from equations~\ref{eq:diskcolumn} and \ref{eq:diskheight}. The time scale is shorter still at the inner edge of the marginally unstable disk, indicating that once the critical mass in the disk has been exceeded, the fragmentation and formation of protostellar clumps is fairly rapid. Consequently, multiple generations of clumps can be perpetually produced and destroyed (through mergers and disruption) during the life time of the O/WR type stars. Observations indicate that most of the O/WR-stars in the central 0.5\,pc are coeval and have formed in a well-defined star formation episode of duration $t_{\rm OWR} \sim6\pm 2$\,Myr \citep{PauGen06}. This time scale then provides a reasonable estimate for the extent of time during which the clumps could have been available for collisions with RGs. Therefore, the hierarchy of time scales is such that $t_{\rm dyn} < t_{\rm cc} \ll t_{\rm lc} < t_{\rm orb} \ll t_{\rm KH} < t_{\rm OWR} \ll  t_{\rm rg} $.

We also tabulate the length of each simulation ($t_{\rm sim}$) and the number of collisions in each simulation ($N_{\rm coll}$). In continuous runs $N_{\rm coll}$ is simply calculated as the number of clump crossings, $t_{\rm sim}/t_{\rm cc}$, whereas in the repeated impact runs it accounts for the integer number of simulated impacts. Note that we keep $N_{\rm coll}$ the same for corresponding continuous and repeated impact runs, so that in both scenarios the RG spends equal time traveling within the clump (approximately $100\,{t_{\rm dyn}}$). This choice allows for a straight forward comparison of results in these two sets of runs later.


\begin{deluxetable*}{ccccccccccc} 
\tabletypesize{\footnotesize} 
\tablecolumns{11} 
\tablewidth{0pt} 
\tablecaption{Simulation results. Run -- simulation name. $M_i$ -- estimated impacted mass. $p_i$ -- estimated impact momentum. $E_i$ -- estimated impact energy. $\langle \Delta p \rangle$ -- average change in momentum per impact. $\langle \Delta E_k \rangle$ -- average change in kinetic energy per impact. $\Delta M_a$ -- estimated mass loss over $100\,t_{\rm dyn}$. $\Delta M$ -- mass loss measured in simulations over $t_{\rm sim}$. $\Delta M/\Delta M_a$ -- ratio of simulated and analytically estimated mass loss. 
\label{tab:massloss}} 
\tablehead{ 
Run  & $M_i$ & $p_i$ & $E_i$ & $\langle \Delta p \rangle$ & $\langle \Delta E_k \rangle$ & $\Delta M_a$ & $\Delta M$ & $\Delta M/\Delta M_a$\\
  &  (M$_{\odot}$)   &  (g\,cm\,s$^{-1}$) &  (erg) &  (g\,cm\,s$^{-1}$) & (erg)  & ($M_\odot$) & ($M_\odot$) & 
} 
\startdata 
A1 &  $7.6 \times 10^{-3}$ & $4.5  \times 10^{38}$ & $6.8 \times 10^{45}$ & $1.9 \times 10^{38}$ & $5.0 \times 10^{45}$ & $3.0\times 10^{-5}$ & $ 9.2 \times 10^{-4}$ & 31 \\ 
RA1 &  $7.6 \times 10^{-3}$   &  $4.5  \times 10^{38}$   & $6.8 \times 10^{45}$ & $ 2.6 \times 10^{38}$ & $7.0 \times 10^{45}$ & $3.0 \times 10^{-5}$ & $9.8 \times 10^{-4}$ &  32 \\ 
B1 &  $7.6 \times 10^{-3}$   &  $4.5  \times 10^{38}$   & $6.8 \times 10^{45}$ & $1.4 \times 10^{38}$ & $4.1 \times 10^{45}$ & $8.6\times 10^{-6}$ & $7.3 \times 10^{-4}$ & 85\\ 
B1\_256 &  $7.6 \times 10^{-3}$   &  $4.5  \times 10^{38}$   & $6.8 \times 10^{45}$ & $1.8 \times 10^{37}$ & $2.7 \times 10^{44}$ & $8.6\times 10^{-6}$ &$6.0 \times 10^{-4}$ & 70\\ 
B1\_300 &  $7.6 \times 10^{-3}$   &  $4.5  \times 10^{38}$   & $6.8 \times 10^{45}$ & $1.9 \times 10^{37}$ & $2.8 \times 10^{44}$ & $8.6\times 10^{-6}$ & $6.3 \times 10^{-4}$ & 73\\ 
B1\_512  & $7.6 \times 10^{-3}$   &  $4.5  \times 10^{38}$   & $6.8 \times 10^{45}$ & $2.6 \times 10^{37}$  & $3.8 \times 10^{44}$ & $3.8\times10^{-6}$ & $3.5 \times 10^{-4}$ & $92$\\ 
A2 & $7.6 \times 10^{-3}$  & $ 9.1  \times 10^{38}$   & $2.7 \times 10^{46}$ & $4.2 \times 10^{38}$ & $2.3 \times 10^{46}$ & $3.6\times 10^{-4}$ & $3.6 \times 10^{-3}$ & 10 \\ 
A3  & $7.6 \times 10^{-3}$  & $1.3  \times 10^{39}$  &  $6.1 \times 10^{46}$ & $ 7.0 \times 10^{38}$ & $5.1 \times 10^{46}$ & $1.6\times 10^{-3}$ & $8.2 \times 10^{-3}$ & 5.1\\ 
A4   &  $7.6 \times 10^{-3}$  & $1.8  \times 10^{39}$  &  $1.0 \times 10^{47}$ & $1.1 \times 10^{39}$  & $1.0 \times 10^{47}$ & $4.4\times 10^{-3}$ & $1.4 \times 10^{-2}$ & 3.2\\ 
A5 &  $7.6 \times 10^{-2}$ & $2.2 \times 10^{39}$ &  $1.7 \times 10^{46}$ & $1.1 \times 10^{39}$ & $1.5 \times 10^{46}$ & $6.1\times 10^{-4}$ & $1.5 \times 10^{-2}$ & 25\\
RA5  &  $7.6 \times 10^{-2}$ & $2.2 \times 10^{39}$ &  $1.7 \times 10^{46}$ & $1.2 \times 10^{39}$  & $1.6 \times 10^{46}$ & $6.1\times 10^{-4}$ &  $1.8 \times 10^{-2}$ & 29\\
A6  & $7.6 \times 10^{-2}$ & $4.5 \times 10^{39}$ &  $6.8 \times 10^{46}$ & $1.4 \times 10^{39}$ & $3.5 \times 10^{46}$ & $7.6\times 10^{-3}$ & $2.9 \times 10^{-2}$ & 3.8\\
A7\_64  & $7.6 \times 10^{-2}$  & $9.1 \times 10^{39}$ & $2.7 \times 10^{47}$ & $3.5 \times 10^{39}$ & $1.5 \times 10^{47}$ & $7.7\times 10^{-2}$ & $1.2 \times 10^{-1}$ & 1.6\\
A7 & $7.6 \times 10^{-2}$  & $9.1 \times 10^{39}$ & $2.7 \times 10^{47}$ & $2.7 \times 10^{39}$ & $1.3 \times 10^{47}$ & $7.7\times 10^{-2}$ &  $8.1 \times 10^{-2}$  & 1.1\\
A7\_256  & $7.6 \times 10^{-2}$  & $9.1 \times 10^{39}$ & $2.7 \times 10^{47}$  & $2.6 \times 10^{39}$ & $1.2 \times 10^{47}$ & $7.7\times 10^{-2}$ & $8.6 \times 10^{-2}$ & 1.1\\
RA7 & $7.6 \times 10^{-2}$  & $9.1 \times 10^{39}$ & $2.7 \times 10^{47}$ & $3.7 \times 10^{39}$ & $1.6 \times 10^{47}$ & $7.7 \times 10^{-2}$ & $1.8 \times 10^{-1}$ & 2.3\\
B7\_300  & $7.6 \times 10^{-2}$  & $9.1 \times 10^{39}$ & $2.7 \times 10^{47}$ & $2.1 \times 10^{39}$  & $1.0 \times 10^{47}$ & $4.5\times 10^{-2}$ & $5.8 \times 10^{-2}$ & 1.3\\
A8    &  $7.6 \times 10^{-2}$  & $1.3 \times 10^{40}$ &  $6.1 \times 10^{47}$ & $4.8 \times 10^{39}$ & $2.3 \times 10^{47}$ & $2.6\times 10^{-1}$ & $1.6 \times 10^{-1}$ & 0.62\\
A9   & $7.6 \times 10^{-2}$  & $1.8 \times 10^{40}$ &  $1.0 \times 10^{48}$ & $5.2 \times 10^{39}$ & $4.4 \times 10^{47}$ & $5.1\times 10^{-1}$ & $2.9 \times 10^{-1}$ & 0.57
\enddata 
\end{deluxetable*}

\section{Results}
\label{sec:results}

\subsection{Energetics of star-clump collisions}
\label{sub:massloss}

Figure~\ref{fig:VisIt_Temp} illustrates the ``star in the wind tunnel" setup used in our simulations. The  3-dimensional snapshots from run A7\_256 show surfaces of constant density (in units of $\rho_{c ,0}$), temperature (in Kelvin), and the Mach number $10\, t_{\rm dyn}$ after the initial collision with the clump. The clump fluid is moving upwards as indicated by the shape of the bow shock that develops as a consequence of the supersonic impact characterized by $\mathcal{M} \simeq 2.0$. The temperature of the flow downstream from the shock raises by a factor of few and is highest directly in front of the star. This hot, high-pressure region is visible as a low density "blister" in the left panel of Figure~\ref{fig:VisIt_Temp}. Also noticeable is the subsonic turbulence that forms behind the star as a consequence of the Kelvin-Helmholtz instability, which is triggered by the velocity shear on the surface of the star.

During each impact a fraction of the RG mass is removed on the account of the star's linear momentum and kinetic energy. To characterize the strength of the encounter and for the purposes of comparisons with the simulations we estimate analytically the amount of mass in the clump impacted by the star, $M_i = \pi \Sigma_c R_*^2$, the impact momentum, $p_i = M_i v_*$, and kinetic energy, $E_i = M_i\, v_*^2 / 2$. We also calculate from simulations the average change in the linear momentum and kinetic energy of the star per impact, $\langle \Delta p \rangle$ and $\langle \Delta E_k \rangle$, after identifying the self-bound mass that represents the star and by measuring the velocity of its center with respect to the background clump fluid. These values are recorded in Table~\ref{tab:massloss}.

We find that the change in linear momentum and kinetic energy of the star measured from simulations is for most runs consistent with the analytic expectations within a factor of few, where the values from simulations tend to be lower than the analytic estimate of the impact momentum and energy. This indicates that not all momentum and kinetic energy available in the impact are used to unbind the stellar envelope or to heat and accelerate the clump fluid around the star. 
 
 The outlier to this trend is the run B1 (a $\Gamma = 4/3$ model) where at higher numerical resolutions the values of $\langle \Delta p \rangle$ and $\langle \Delta E_k \rangle$ fall short of the analytic estimates by a factor of about 20. This is an encounter at a relatively low velocity of $300\,{\rm km\,s^{-1}}$, where only a small fraction of the total impact momentum and energy is sufficient to unbind a low density outer envelope extending beyond $0.8R_*$ while the more compact remainder of the star continues on its trajectory like a bullet. As a consequence, the star does not experience a big change in $\langle \Delta p \rangle$ and $\langle \Delta E_k \rangle$. On the other hand, in more energetic collisions captured in simulation B7\_300, which affect deeper layers of the stellar envelope, $\langle \Delta p \rangle$ and $\langle \Delta E_k \rangle$ approach the total impact momentum and kinetic energy within a factor of a few. It follows that the strength of star-disk interactions, besides the collision velocity and column density of the clump, also depends on the structure of the star, the fact that is not captured in the simple analytic estimates of $p_i$ and $E_i$. 

\subsection{Mass loss and the structure of the star}
\label{sub:massloss}

\begin{figure*}[!t] 
\epsscale{1.0}  
\plottwo{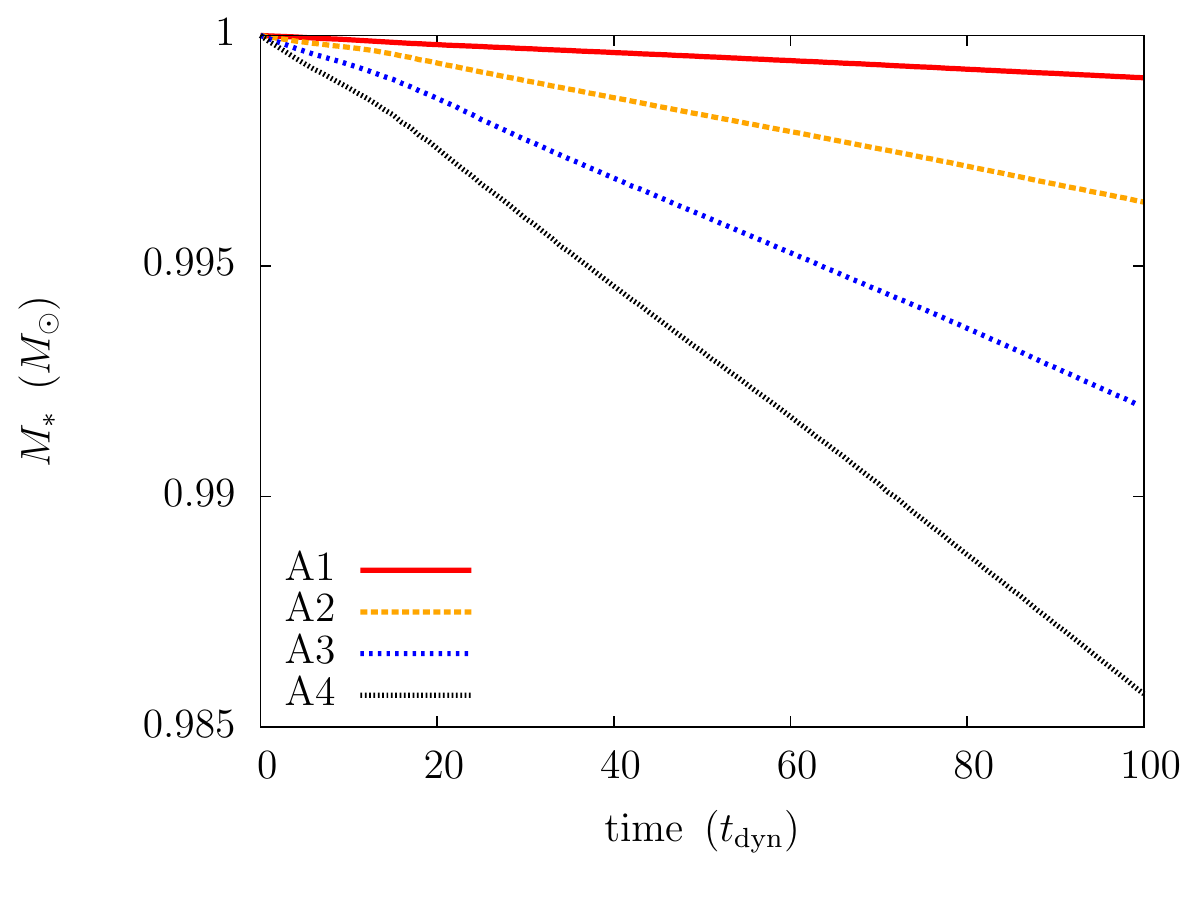}{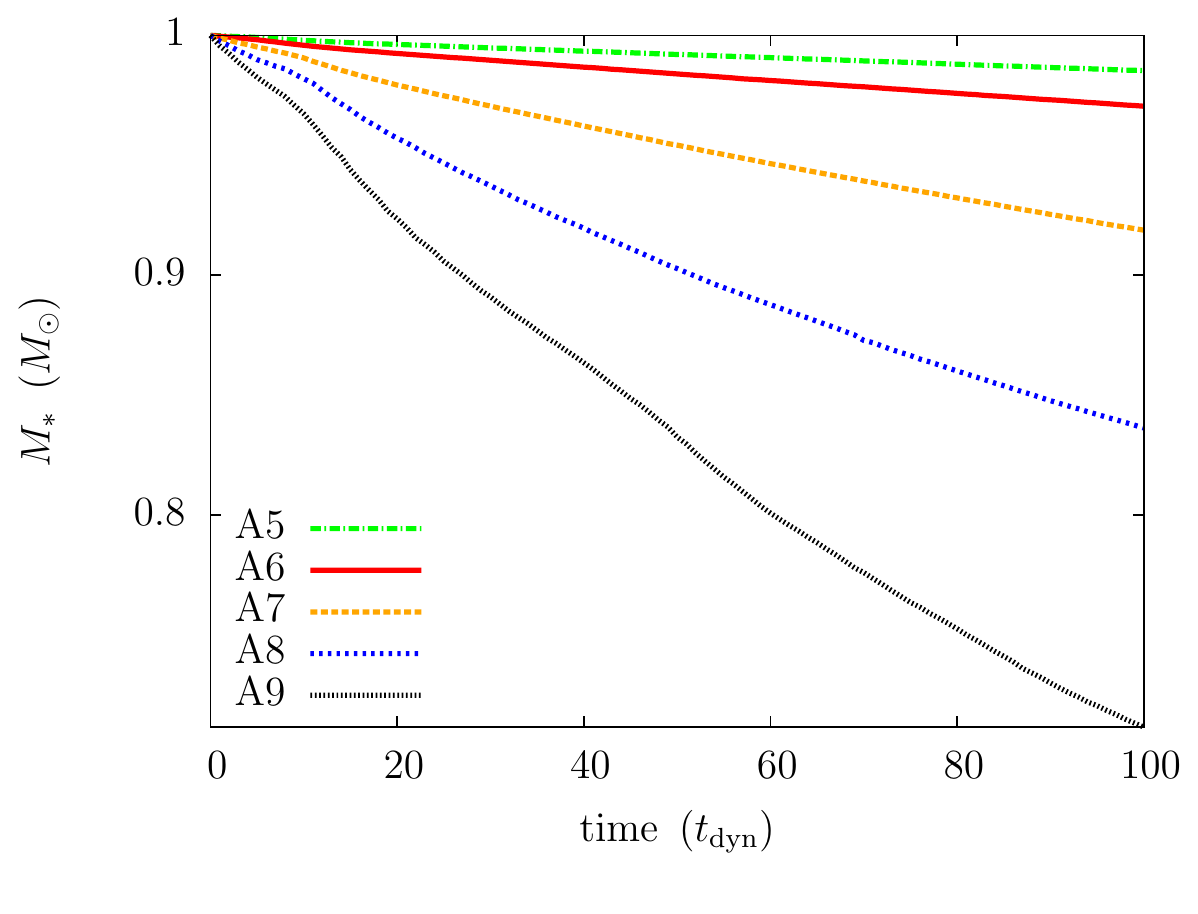}
\caption{Total mass of RG as a function of time in the encounters with clumps of column density $10^7\,{\rm g\,cm^{-2}}$ (left) and $10^8\,{\rm g\,cm^{-2}}$ (right). Different line styles correspond to the following initial impact velocities: 1200 (dotted black), 900 (dotted blue), 600 (dashed orange), 300 (solid red), and 150\,${\rm km\,s^{-1}}$ (dash-dot green). Note the difference in scale of the $y$-axis in the two panels.}
\label{fig:MassLoss}
\end{figure*}

Figure~\ref{fig:MassLoss} and entries in Table~\ref{tab:massloss} illustrate the change in the total mass of the RG measured from simulations A1$-$A9 over a period of $100\,t_{\rm dyn}$. Note that the period of $100\,t_{\rm dyn}$ represents multiple clump crossing times (in the "continuous" setup) and for simulations described here corresponds to a number of collisions  $N_{\rm coll} = 2 -25$. The mass of the RG in simulations is calculated at every time step by evaluating the gravitationally bound mass of the star.  For a range of simulated impact velocities the RG loses $\lesssim$1.5\% of its mass in collisions with a clump with column density of $10^7\,{\rm g\,cm^{-2}}$ (left panel of Figure~\ref{fig:MassLoss}). The mass loss is more substantial and reaches up to 30\% for collisions with the clump of column density $10^8\,{\rm g\,cm^{-2}}$. 

Because the values of $\Delta M$ reported in Table~\ref{tab:massloss} are quite low in some runs, we ensure that the smallest mass loss that is reliably measured in our simulations is several orders of magnitude lower. Along similar lines, our resolution studies indicate that the mass loss measurements in $128^3$ runs are numerically converged (see Appendix for more information). For example, a comparison of run A7 with A7\_256 yields a range, $\Delta M = (8.1 \pm 0.5)\times 10^{-2}\,M_\odot$, which corresponds to an effective error of about 6\%.

In Table~\ref{tab:massloss} we compare the mass loss measured in simulations to the analytic estimate, $\Delta M_a$, calculated following the approach by \citet{ArmZur96}. This calculation is based on the analytic theory developed by \cite{1975ApJ...200..145W}, who considered the effect of a supernova blast wave on the companion in a binary star system. For the purpose of the analytic estimate we assume that momentum transfer is the dominant mechanism responsible for the stripping of the RG and neglect the mass loss by ablation (shock heating of the material on the surface of the star). This is justified because ablation becomes important at high Mach numbers characteristic of supernova blast waves, when the blast wave collision velocity significantly exceeds the escape velocity at the surface of the impacted star. 

We calculate the critical radius, $R_{\rm crit}$, where the momentum transferred to a cylindrical shell is sufficient to accelerate it to the stellar escape velocity at that radius:
\begin{equation}
\Sigma_* (R) v_{\text{es}} (R) - v_* \Sigma_c = 0\, .
\label{eq_sigma}
\end{equation}
Here $v_{\text{es}} (R)$ and $\Sigma_* (R)$ are the escape velocity and the column density of the RG at the radius $R \leq R_*$, respectively. The mass loss $\Delta M_a$ then corresponds to the mass in the envelope exterior to a cylinder of radius $R_{\rm crit}$.

\begin{figure*}[!t] 
\epsscale{1.0}  
\plottwo{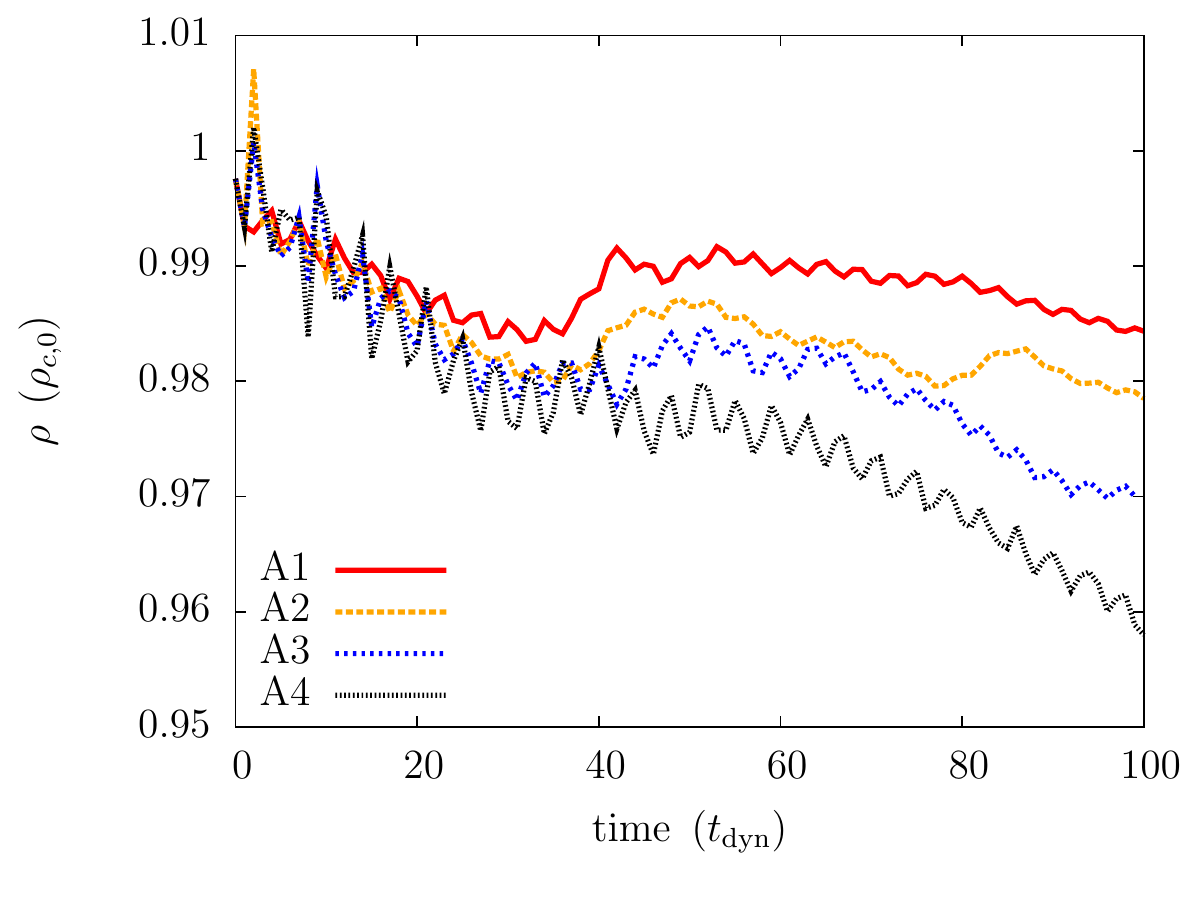}{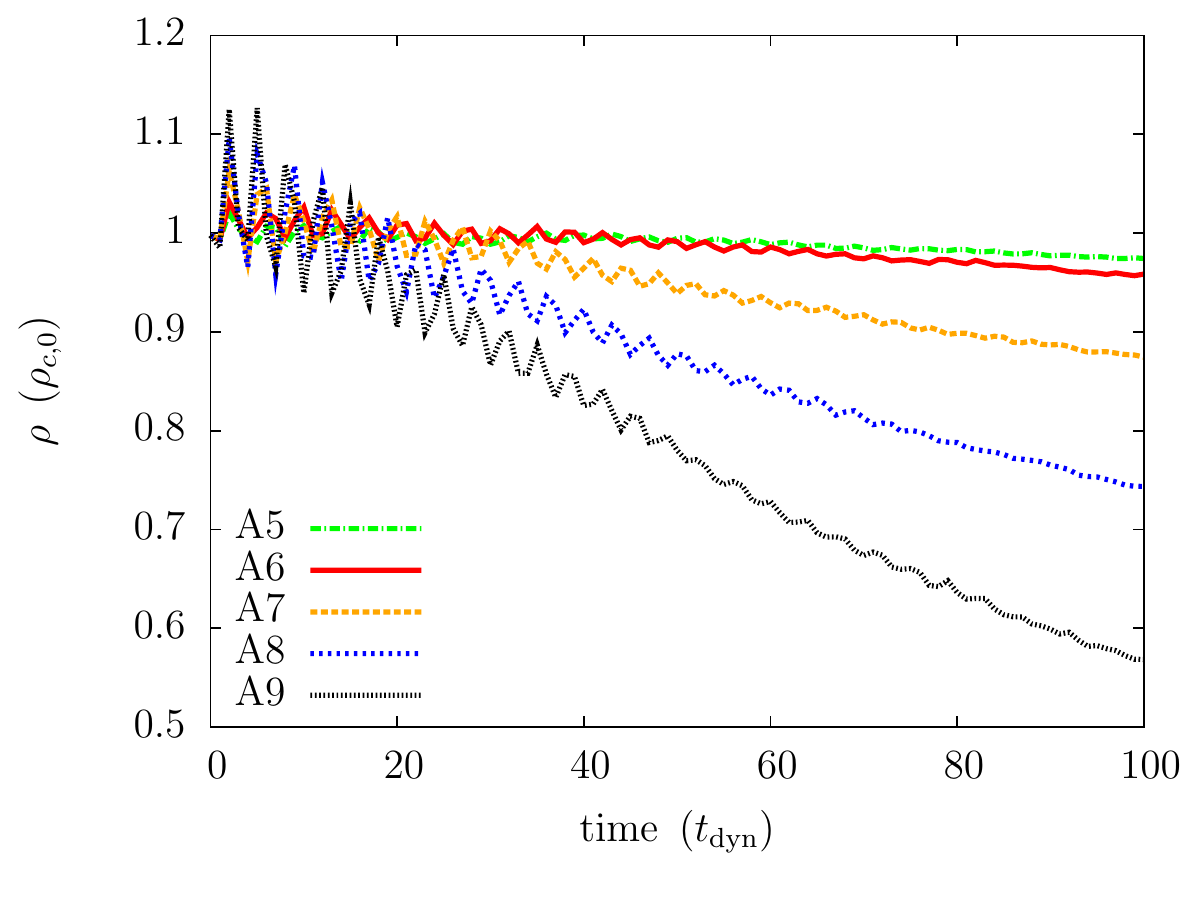}
\caption{Central density of RG as a function of time in the encounters with clumps of column density $10^7\,{\rm g\,cm^{-2}}$  (left) and $10^8\,{\rm g\,cm^{-2}}$ (right). Line styles represent different velocities, as in Figure~\ref{fig:MassLoss}. Note the difference in scale of the $y$-axis in the two panels.}
\label{fig:CentralDensity}
\end{figure*} 

\begin{figure}[!t] 
\epsscale{1.1}  
\plotone{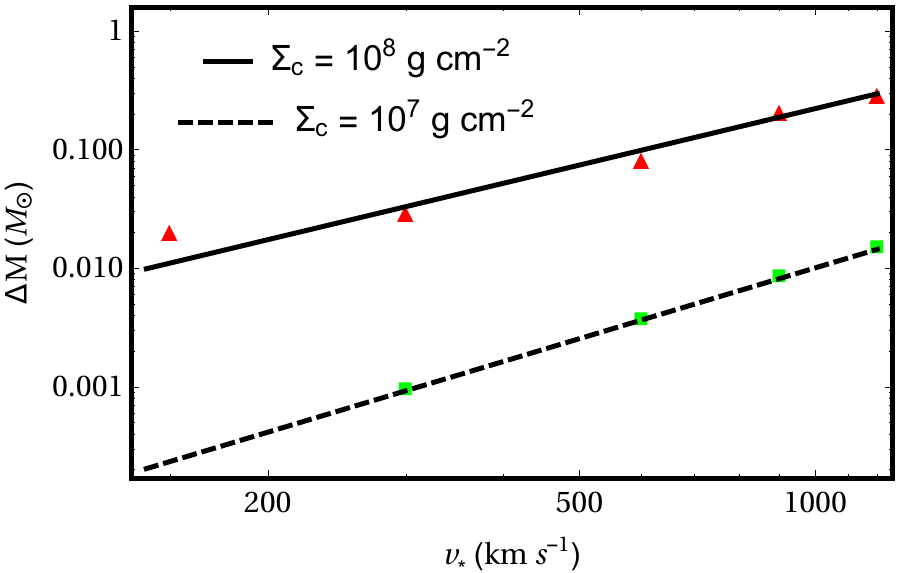}
\caption{Mass loss as a function of the impact velocity for clump column densities $\Sigma_c = 10^7$ (green squares) and $10^8$ g cm$^{-2}$ (red triangles), all measured over a period of $100\,t_{\rm dyn}$. Solid and dashed lines are the best fit to the data given by the least squares regression.} 
\label{fig:VelocityVSMassLoss}
\end{figure} 

We find that in most cases (except A8 and A9) the analytic estimate tends to underestimate the mass loss relative to the simulations. Furthermore, $\Delta M_a$ provides a better estimate of mass loss (within a factor of few) for runs with $N_{\rm coll} > 3$, in which stars tend to slow down due to the loss in kinetic energy. In these cases the mass loss averaged over many clump crossings more closely corresponds to the analytic value while it tends to be higher than average during the initial couple of passages (see discussion in Section~\ref{S_repeated}). 

Figure~\ref{fig:CentralDensity} shows the central density of the RG as a function of time for the same set of runs shown in Figure~\ref{fig:MassLoss}. They illustrate the perturbation in the inner, core region of the star and its departure from initial equilibrium as a consequence of collision. The perturbed star "rings" as illustrated by the presence of damped oscillations in the central density with characteristic time scale of $\sim 3\,t_{\rm dyn}$. In addition to the core the outer layers of the star also oscillate on longer time scales, an effect noticeable in the first $\sim 40\,t_{\rm dyn}$ for collisions with lower density clumps (left panel of Figure~\ref{fig:CentralDensity}). An RG that enters a denser clump (right panel) is stripped of its outer envelope and is also confined by the stronger pressure of the surrounding medium. Consequently, its central density oscillations decay more rapidly and oscillations of the outer envelope are absent. At the same time, the RG tunneling through the $10^8\,{\rm g\,cm^{-2}}$ medium experiences a precipitous drop in the central density as its mass decreases and the star expands to adjust to a new hydrostatic equilibrium. 

\begin{figure*}[t]
\centering
\includegraphics[scale=0.14]{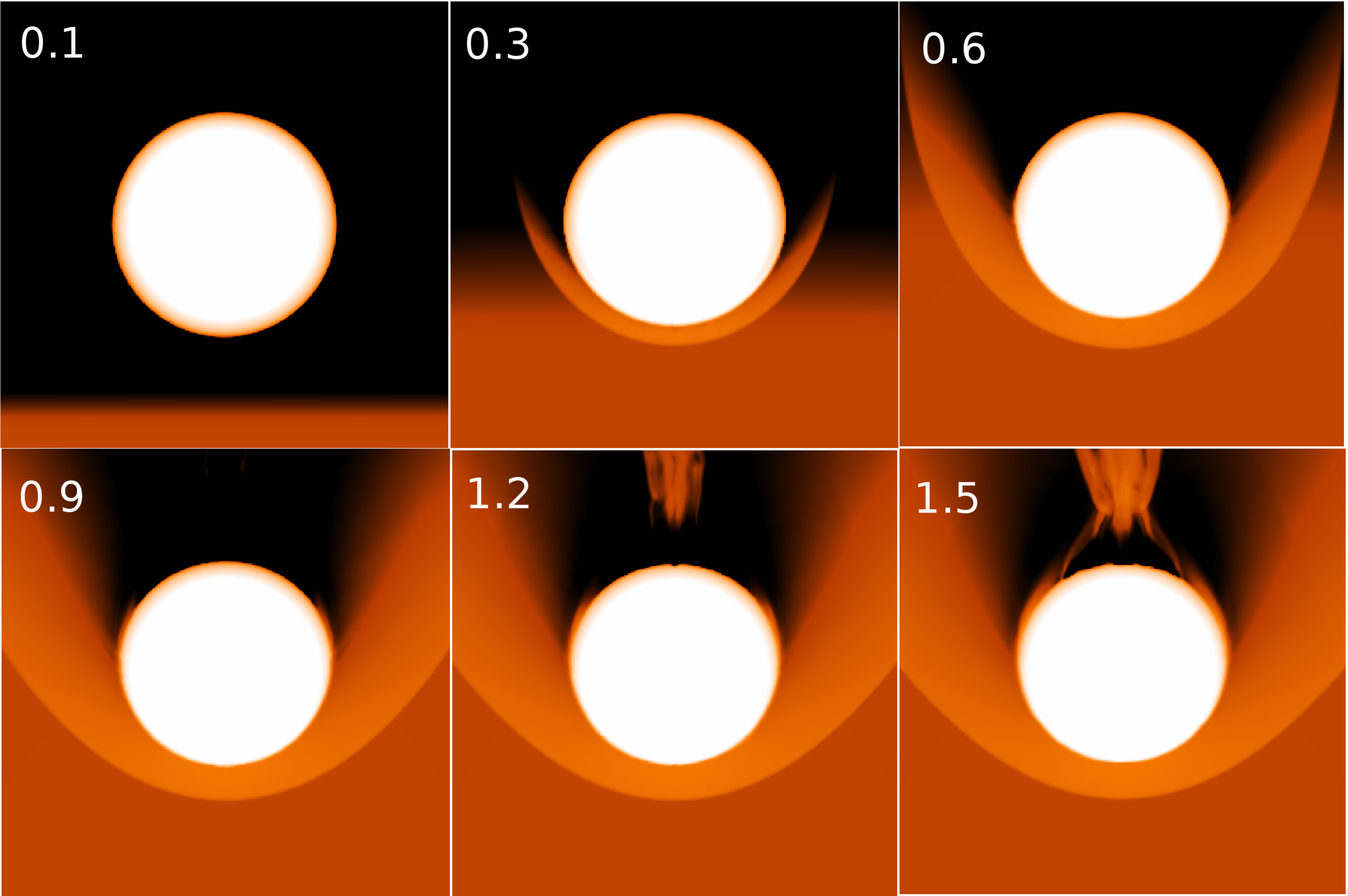} 
\includegraphics[scale=0.14]{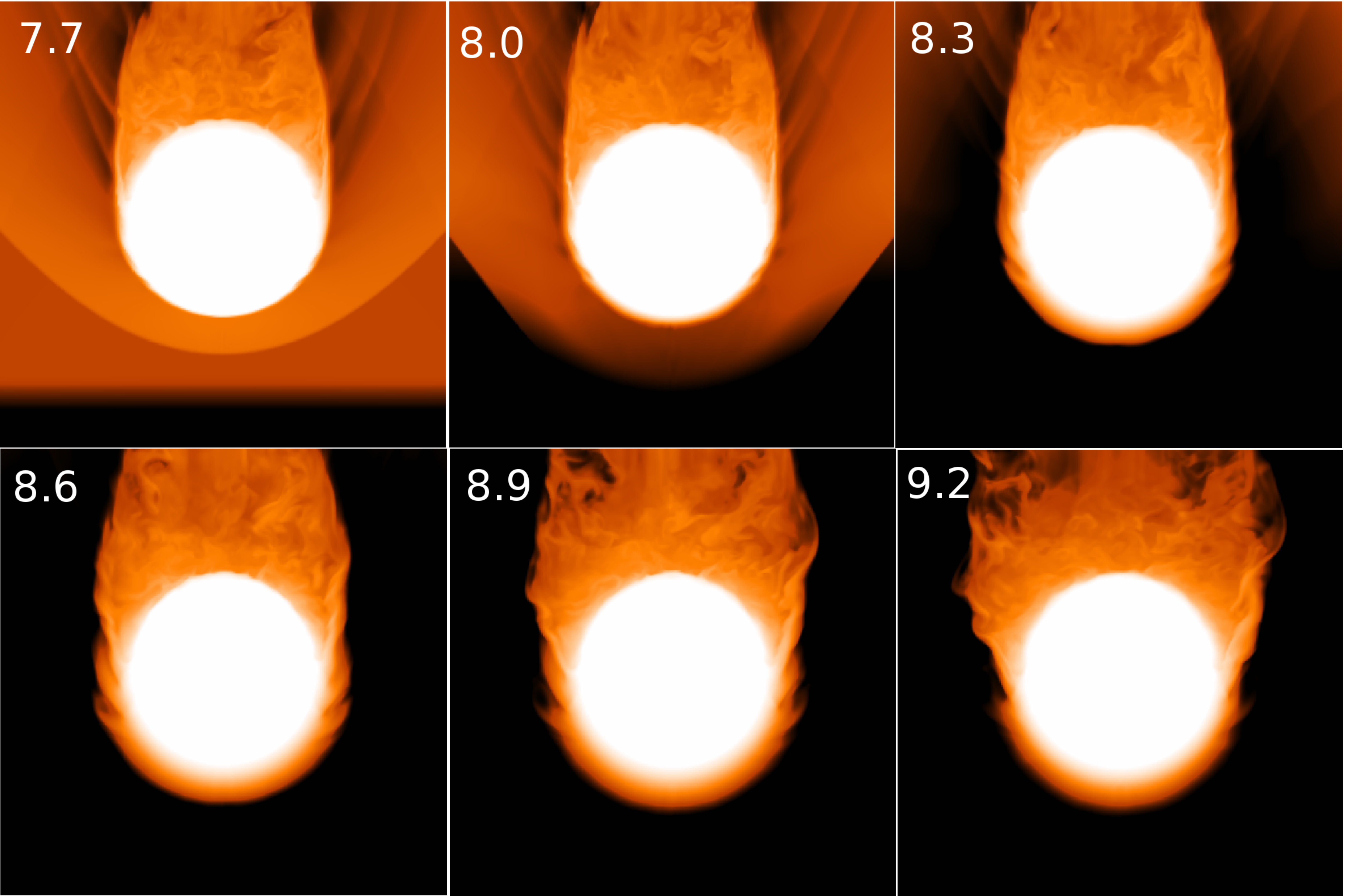} 
\includegraphics[scale=0.35]{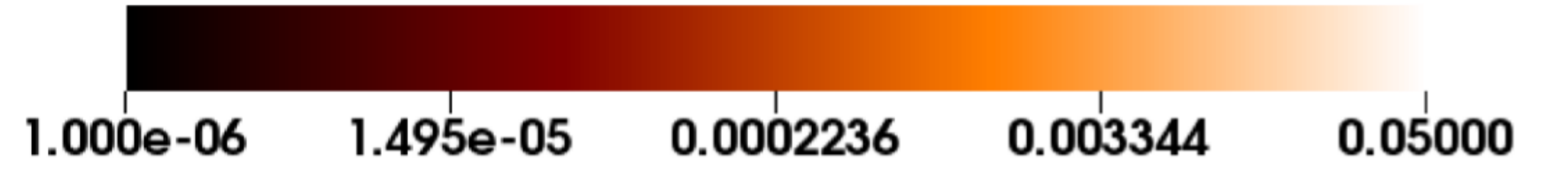}
\caption{Sequence of two-dimensional snapshots showing RG as it impacts (left) and exits (right) the clump in RA7\_256. Each snapshot is a slice through the center of the star. Color bar indicates density in units of $\rho_{c,0}$ and time stamps indicate the time each snapshot was taken in the simulation (in units of $t_{\rm dyn}$).}
\label{fig:EnterAndExiting}
\end{figure*}

A trend observed in all runs with continuous setup is that the mass loss increases with the increasing impact velocity as shown in Figure~\ref{fig:VelocityVSMassLoss}. For example, in collisions with the lower clump density the impact velocity triples from run A1 to A3 and the measured increase in mass loss is by a factor of 9. Similarly, the velocity doubles from run A1 to A2 and from A2 to A4 and in both cases the corresponding increase in mass loss is by a factor close to 4, indicating the mass loss dependance $\propto v_*^2$. This proportionality follows from equation~\ref{eq_sigma} where the total column density encountered by the star traveling through the continuous medium of density $\rho_{\rm med}$ is determined by $\Sigma_c =  \rho_{\rm med}\, v_*\, t_{\rm sim}$ and hence, $ \Sigma_* (R) \propto \Delta M \propto v_*^2$. 

In collisions with the higher density clump, the impact velocity triples from the run A6 to A8 while the measured increase in mass loss is a factor of $\sim7$. In a sequence of runs from A5, A6, A7 to A9 the mass loss increases by an average factor of 2.6 each time the impact velocity doubles, indicating the dependance on $v_*$ which is super-linear but not quite quadratic. This is due to the fact the star is slowing down as it travels through the clump, an effect which is more pronounced at the higher clump density. 

These trends are confirmed by the fits to the data points plotted in Figure~\ref{fig:VelocityVSMassLoss}, which indicate different exponential dependance on velocity for collisions with the lower and higher density clumps
\begin{eqnarray}
\log \Delta M_7 &=& -7.96 + 1.99 \log v_*\\
\log \Delta M_8 &=& -4.74 + 1.35 \log v_*
\end{eqnarray}
The linear fits are obtained using the basic least squares regression, where $\Delta M$ is in units of solar masses and $v_*$ is in km\,s$^{-1}$. The subscripts on $\Delta M$ indicate the fits to the $\Sigma_c=10^7$ and  $10^8\,{\rm g\,cm^{-2}}$ data points.

In addition to the A-runs where stars are modeled as $\Gamma=5/3$ polytropes we also carried out simulations with $\Gamma=4/3$ polytropes (B-runs). These include runs B1\_300 and B7\_300, which are counterparts to the A1 and A7, respectively, in terms of the impact velocity and clump column density. In both cases the measured mass loss for the A and B models was comparable to within about 30\%, with B runs resulting in a smaller mass loss. In comparison to the $\Gamma=5/3$ polytropes, $\Gamma=4/3$ models are vulnerable to stripping to a much smaller radius within the star (Figure~\ref{fig:AdiabaticIndex}). Their tenuous and extended envelopes however contain only a small fraction of the total mass and the combination of these two effects results in a smaller overall mass loss for $\Gamma=4/3$ polytropes. We therefore consider the value of the mass loss measured from the A-runs as an upper bound on the mass loss from both models.

The discussion of mass loss up to this point focused on the runs with the continuous setup. In the following section, we discuss the effect of repeated impacts on the mass loss and structure of the star.

\subsection{The effect of repeated impacts}
\label{S_repeated}

During realistic encounters with a fragmenting disk an RG may have multiple collisions with individual clumps. It has been pointed out by \citet{ArmZur96} that such successive impacts can be very efficient in removing the outer layers of the RG star. To investigate this effect we carried out the runs RA1, RA5, and RA7, which are counterparts to A1, A5, and A7, respectively. In RA runs, the RG travels through the clump and exits after an interval of time $t_{\rm cc}$, then continues to move through the low density medium for $2t_{\rm cc}$ before collision with another clump. The choice for the interval of time between collisions in our simulations is fiducial and in all cases amounts to $> 16\,t_{\rm dyn}$, allowing the star to relax before the next impact. In reality, the time scale between repeated RG collisions depends on the number of clumps and orientation of the RG orbit relative to the fragmenting disk, but is always expected to be considerably longer than $\sim t_{\rm dyn}$ (see \S~\ref{sec:Discussion} for discussion). 


\begin{figure}[t]
\includegraphics[trim= 0 0 0 -30, scale=0.92]{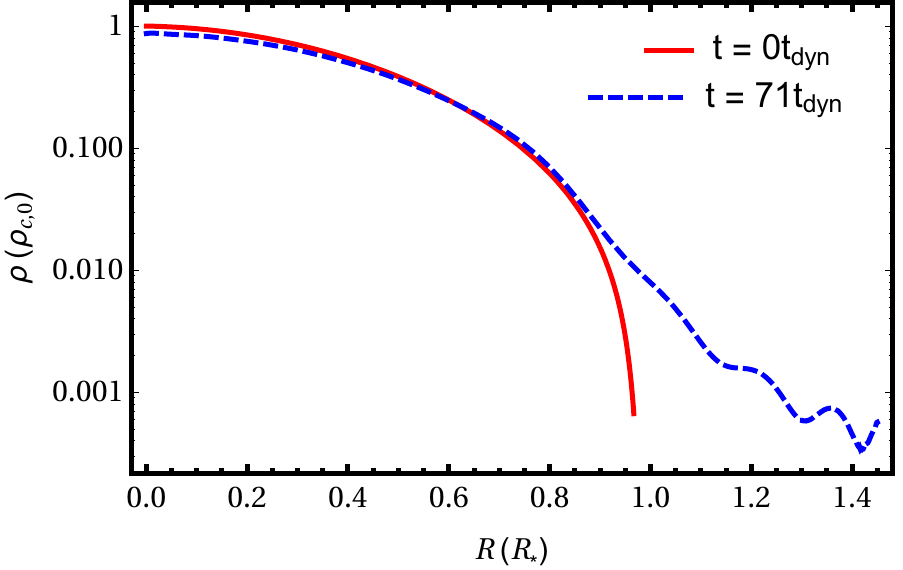} 
\caption{Spherically averaged density profiles of RG in RA7 at $t=0$ (red, solid line) and as it emerges from a clump after the third collision at $t= 71\,t_{\rm dyn}$ (blue, dashed). Expansion of the envelope increases the cross section of the star to subsequent collisions, making the mass loss efficient.}
\label{fig:envelope}
\end{figure}

\begin{figure*}[!t] 
\epsscale{1.0}  
\plottwo{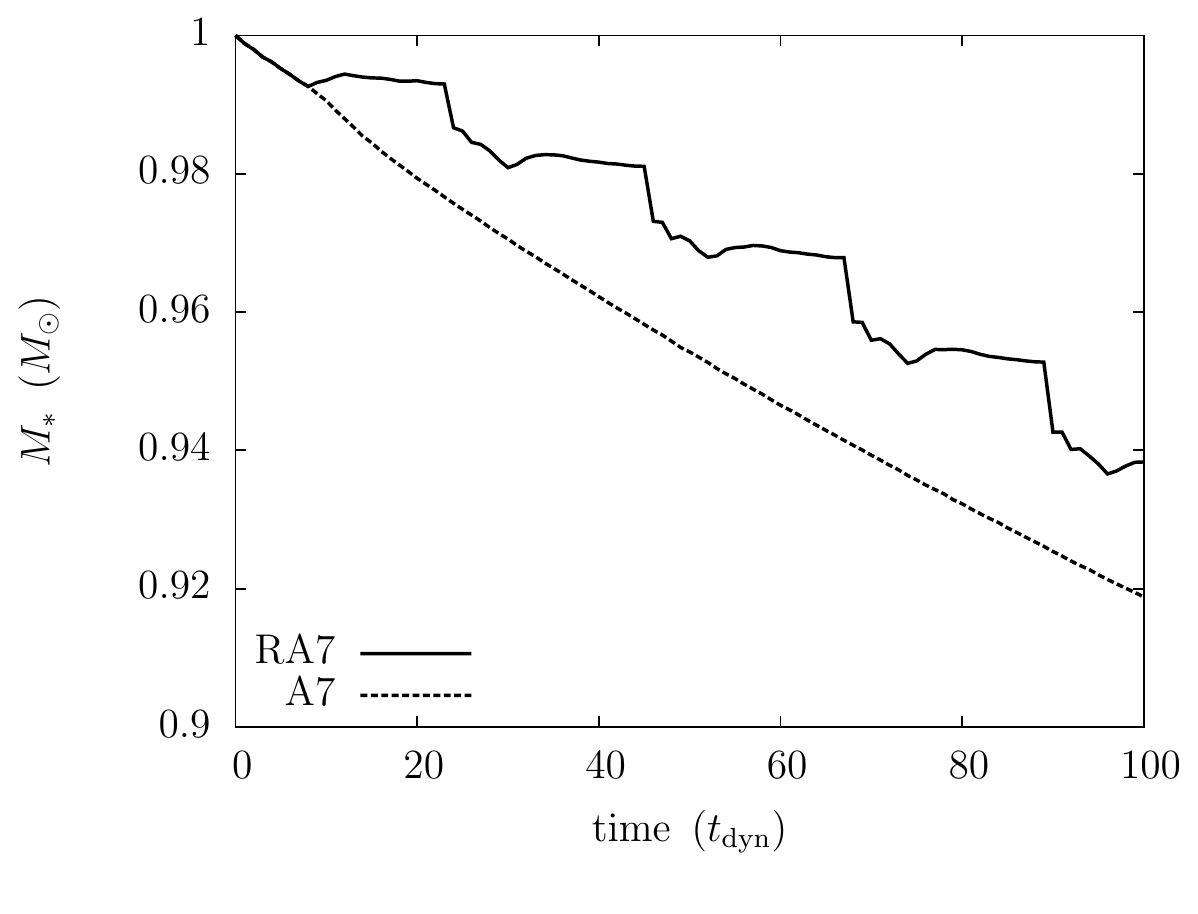}{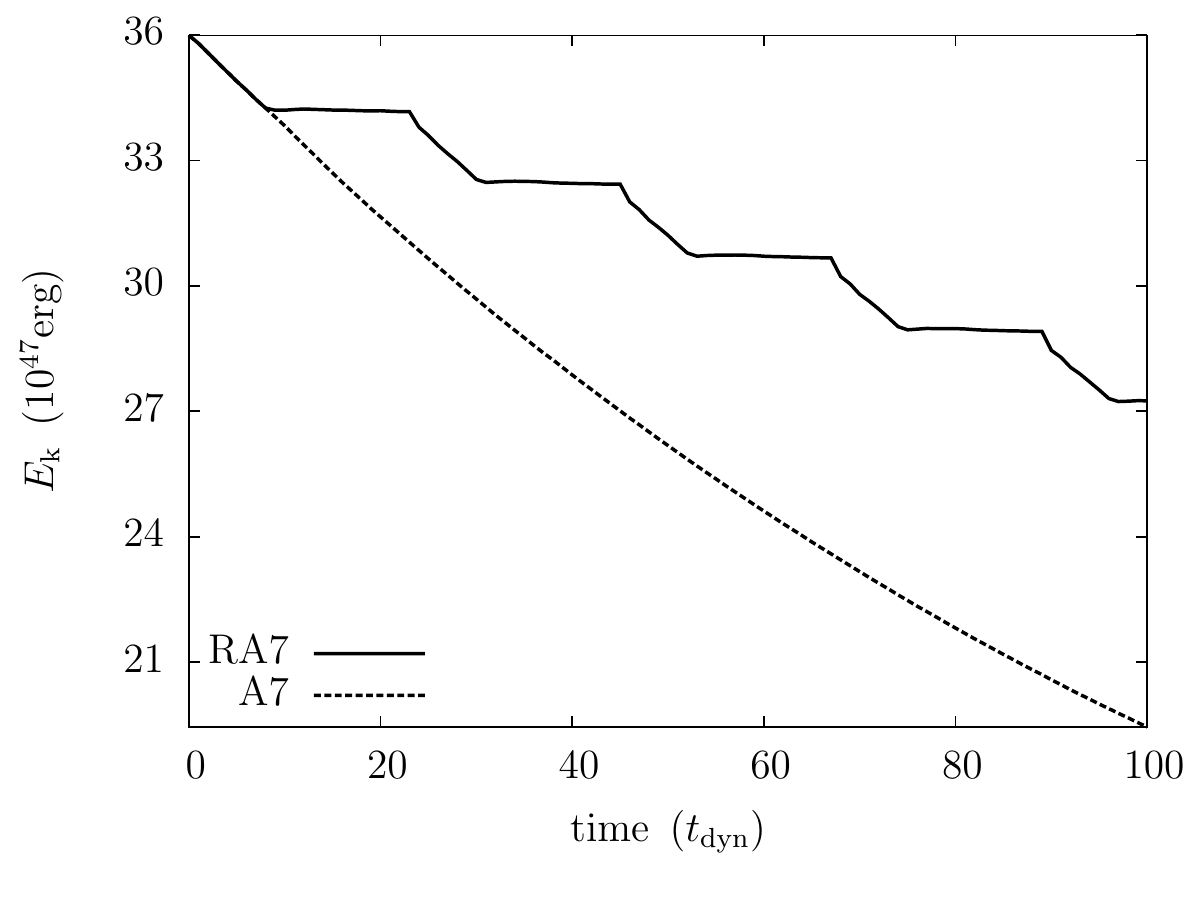}
\caption{Total mass (left) and kinetic energy (right) of the star as a function of time for the runs A7 (dotted) and RA7 (solid line). The step like appearance of the solid curves illustrates the star losing mass and kinetic energy every time it collides with a clump. The clump crossing time for RA7 is $t_{cc} = 8\,t_{\rm dyn}$ and time between the collisions was arbitrarily set to $2t_{\rm cc}$.} 
\label{fig:MassLoss_periodic}
\end{figure*} 

In all RA runs we assume that successive collisions involve clumps of the same column density. A visualization of this is provided as a sequence of snapshots from run RA7\_265 shown in Figure~\ref{fig:EnterAndExiting}. The left set of panels shows the density of the surrounding medium as the star enters the clump for the first time. The characteristic bow shock in front of the star and the plume of gas stripped from the star develop quickly and within only a few $\,t_{\rm dyn}$ after the impact. 

The RG whose envelope is truncated through stripping tends to expand beyond its original size in order to achieve a new hydrodynamic equilibrium and the central density of the star decreases. The smooth evolution through a sequence of equilibria is interrupted once RG exits the clump and finds itself almost instantaneously in a low pressure environment of the interstellar medium. Figure~\ref{fig:envelope} shows the spherically averaged density profiles of the star in the RA7 run at $t=0$ and as it emerges from the clump at  $t=71\,t_{\rm dyn}$, in the aftermath of its third successive impact. The latter profile illustrates subsequent expansion of the outer layers of the star, which decreases their binding energy and increases the cross section for collision with the next clump, making RG more susceptible to mass loss. The envelope expansion is also noticeable in the right panels of Figure~\ref{fig:EnterAndExiting} as a diffuse atmosphere that forms in front of the star after it leaves the clump.

Figure~\ref{fig:MassLoss_periodic} shows the total mass (left) and kinetic energy (righ panel) of RG for runs A7 and RA7. The left panel illustrates that RA7 has five star-clump encounters during the initial $100\,t_{\rm dyn}$ and each time RG impacts a clump there is a steep drop in the mass of the star. When RG exits the clump, some of the low density atmosphere and trailing gas retracts to the star, increasing its mass by a small amount. Figure~\ref{fig:MassLoss_periodic} shows RG losing more mass in run A7 than run RA7, because in A7 the star tunnels through a larger column of gas during the $100\,t_{\rm dyn}$ interval (equivalent to 12 star-clump collisions). Comparing the mass loss in A7 and RA7 after the star has completed 12 encounters in both runs, we find that in the RA7 run RG loses two times more mass than in A7 (18\% versus 8.1\%, respectively). Similar trend is noticeable for run pairs A1 -- RA1 and A5 -- RA5, although less pronounced given the fewer simulated impacts.

Comparing the mass loss measured in repeated impact simulations to the analytic estimates we find that the latter underestimate it by a factor of $\sim{\rm few}-30$ (last column of Table~\ref{tab:massloss}).  Also, the simulated values of mass loss in repeated impact scenarios tends to diverge slightly more from the analytic estimates than the continuous setup runs. This points to the importance of repeated impacts for the evolving structure of the star which is not captured by the analytic treatment.

We also find in all runs with repeated impacts that the loss of kinetic energy is larger relative to the continuous setup runs, and so is the loss of linear momentum. The right panel of Figure~\ref{fig:MassLoss_periodic} illustrates the evolution in the kinetic energy of a star as a function of time for runs A7 and RA7. In both runs the mass loss and kinetic energy largely mirror one another, indicating that the surface layers of the star are unbound on account of its kinetic energy. After a dozen impacts we measure the change in the kinetic energy of the star $\Delta E_k^{\rm tot} = 1.9 \times 10^{48}\,{\rm erg}$ in RA7 and $\Delta E_k^{\rm tot} = 1.6 \times 10^{48}\,{\rm erg}$ in A7\footnote{We tabulate the average change in the kinetic energy {\it per impact} in Table~\ref{tab:massloss}.}. This degree of kinetic energy loss  is larger than the binding energy of the star, $E_g \sim GM_*^2/R_* = 3.8\times 10^{47}\,{\rm erg}$ and it leads to a slowdown of the star to about a half of the initial velocity. This indicates that damage to the star can in principle be substantial. We indeed measure an RG mass loss of order of tens of percent of the initial mass after 12 impacts. The fact that the star is not completely destroyed even though $\Delta E_k^{\rm tot} > E_g$ suggests that the remaining fraction of the RG's kinetic energy is spent on accelerating and heating the clump gas. 

\subsection{The value of the mass loss factor, $f_{\rm loss}$}

In order to provide a straight forward way to estimate the mass loss for an arbitrary number of impacts we use our simulations to calibrate the mass loss factor $f_{\rm loss}$, defined by ASC as the ratio of mass losses for two consecutive impacts.
\begin{equation}
f_{\rm loss} = \frac{ \Delta M_{i + 1} }{ \Delta M_i } 
\end{equation} 
where $f_{\rm loss}>1$ and $\Delta M_i$ is the mass loss after the $i$th impact.  Assuming that $f_{\rm loss}$ remains relatively constant from one impact to another implies that the $(i+1)$th impact strips a mass $\Delta M_{i+1} = f_{\rm loss}^{i} \Delta M_1$. Summing over $n$ impacts yields the total mass loss
\begin{equation}
\Delta M = \Delta M_1 \sum_{i=0}^{n-1} f_{\rm loss}^{i} = \Delta M_1 \frac{f_{\rm loss}^n - 1}{f_{\rm loss} - 1}
\end{equation}
ASC note that the typical value of $f_{\rm loss}\approx 2$ for RGs with radii $R_* = 150\,R_\odot$ \citep[similar to those simulated by][]{ArmZur96} and $f_{\rm loss} < 2$ for smaller RGs. For their analytic estimate ASC adopt $f_{\rm loss} =1.01-1.1$. Although it is not obvious how was this range of values selected, it turns out to be a rather good assumption for $R_* = 10\,R_\odot$ RGs.

\begin{deluxetable}{cccccccc} 
\tabletypesize{\footnotesize} 
\tablecolumns{9} 
\tablewidth{0pt} 
\tablecaption{Values of the mass loss factor \label{tab:floss}} 
\tablehead{ 
Run  & $f_1$ & $f_2$ & $f_3$ & $f_4$ & $f_5$ & $f_6$ & $\langle f_{\rm loss} \rangle$ 
} 
\startdata 
A1 &  0.97 & - & - & - & - & - & 0.97 \\
RA1 & 1.34 & - & - & - & - & - & 1.34 \\
B1 & 0.87 & - & - & - & - & - & 0.87 \\
A2 & 1.16 & 0.99 & - & - &  -  &  - & 1.07 \\
A3 & 1.32 & 0.93 & 0.99 & 1.02 & - & - & 1.06 \\
A4 & 1.27 & 1.02 & 0.95 & 1.00 & 1.05 & 0.99 & 1.04 \\
A5 & 0.82 & 0.97  & - & - & - & - & 0.89 \\
RA5 & 1.21 & 1.02  & - & - & - & - & 1.11  \\
A6 & 0.73 & 0.97 & 0.97 & 0.96 & 1.00 & - & 0.92 \\
A7 & 1.26  & 0.81 & 0.91 & 0.94 & 0.97 & 0.96 & 0.97 \\
RA7 & 1.69 & 1.10 & 1.14 & 1.07 & 0.97 & 1.02 & 1.10 \\ 
B7 & 1.29 & 0.85 & 0.99 & 0.94 & 0.97 & 0.98 & 0.99 \\
A8 & 0.78 & 1.60 & 0.93 & 0.86 & 0.93 & 0.96 & 0.99 \\
A9 & 0.79 & 1.44 & 1.13 & 0.86 & 0.87 & 0.92 & 1.00 
\enddata 
\end{deluxetable}

Table~\ref{tab:floss} shows the values of the mass loss factor measured from simulations for consecutive impacts, $f_i$, and their average value, $\langle f_{\rm loss}\rangle$. We report the mass loss factors for the first seven impacts only, where $f_i= \Delta M_{i+1}/\Delta M_i$. Note that seven impacts were arbitrarily chosen for illustration and that some simulations had more and some fewer impacts (in such cases Table~\ref{tab:floss} shows no values). $\langle f_{\rm loss}\rangle$ was however calculated as an average of all impacts captured in a given run, even if they are not listed in this table.

 For majority of runs the value of $\langle f_{\rm loss}\rangle$ is close to unity and in the range $(0.9-1.1)$ for runs with three or more impacts. The simulations show that $f_i$ fluctuates significantly during the first few impacts and then asymptotes to a value close to 1. The RA runs are characterized by values of $\langle f_{\rm loss}\rangle > 1$, reflecting more efficient stripping measured for repeated impacts. In all simulations the first two to three impacts also seem to be the most damaging to the star. 
 
Note that the calculated values of $\langle f_{\rm loss}\rangle$ apply to a finite number of star-clump collisions. However, for stars that experience many more energetic collisions, the star may eventually slow down to the degree when it can be captured by a massive clump or lose so much mass that all of its envelope is effectively disrupted. We do not explore these regimes in our simulations. 
  

\section{Discussion}
\label{sec:Discussion}

\subsection{The star-clump collision time scale}
\label{S_tcoll}

Earlier sections quantify physical conditions necessary to remove a significant fraction of mass ($\gtrsim 10$\%) from RGs similar to those commonly found in the central parsecs of the GC. We find that this is possible for stars that experience tens of collisions with compact, high density clumps ($\gtrsim 10^{8}$ g cm$^{-2}$) that form in late stages of disk fragmentation. The relevant question is then, how likely is it that a star will experience multiple collisions with such clumps while in RG or HB phase of its life?

Assuming that RGs orbiting in the GC cross the plane of the fragmenting disk twice per orbit, we estimate the rate of collisions from geometric arguments as
\begin{equation}  
\Gamma_{\rm coll} = \frac{2N_c}{t_{\rm orb}}\left(\frac{R_c}{R_d}\right)^2
\end{equation}
where $N_c$ is the total number of clumps in the disk and $R_d$ is the radius of the fragmenting disk. One can then estimate the average time between two consecutive star-disk collisions as
\begin{eqnarray}
\label{eq_tcoll}
&& t_{\rm coll}  =  \frac{1}{\Gamma_{\rm coll}} \approx \\
& & 8\times 10^{9}\,{\rm yr} 
\left( \frac{N_c}{100} \right)^{-1} 
\left( \frac{R_d}{0.5\,{\rm pc}} \right)^2
\left( \frac{R_c}{10^{-5}\,{\rm pc}}  \right)^{-2}
\left( \frac{r}{0.1\,{\rm pc}} \right)^{3/2} \nonumber 
\end{eqnarray}
This is significantly longer than both the RG / HB stellar phase for a $1\,M_\odot$ star ($t_{\rm rg}\sim 10^8\,{\rm yr}$) and the time scale over which clumps may have been present in the GC (constrained by the life time of the O/WR type stars, $t_{\rm OWR} \sim6\pm 2$\,Myr), indicating that collisions are unlikely. However, there are several additional factors that could affect this time scale by several orders of magnitude.

\begin{itemize}
\item The value of $N_c$ shown in equation~\ref{eq_tcoll} is motivated by observations of the GC that show existence of a young stellar disk containing about 100 WR/O stars \citep{PauGen06, LuGhe09, bartko09} and by theoretical models which indicate that the mass of the fragmenting disk must have been at least ${\rm few}\times 10^4\,M_\odot$ \citep{levin07}. There are indications however that the observed system of stars is the remnant of what used to be a more densely populated stellar disk created in a common star formation event \citep{yelda14}. If so, the number of clumps could have been significantly higher in the past, given that only a fraction of the clumps would have formed stars, and only a fraction of the stars remains in the observed disk.
 
\item In our simulations we assume that all clumps have the same mass, $M_c = 100\,M_\odot$ (see Section~\ref{sub:runs}). In that case the dependance of $t_{\rm coll}\propto R_c^{-2} \propto (\Sigma_c/M_c)^2$ implies that collisions of RGs with even more compact clumps ($\Sigma_c > 10^8\,{\rm g\,cm^{-2}}$) are increasingly unlikely because of their diminished cross-section. More realistically however, the clumps will be characterized by some mass distribution function and should the average clump be more massive than $100\,M_\odot$, this could further increase the chance for repeated collisions.   

\item Furthermore, the RG orbit may intersect with the fragmenting disk at some oblique angle and be driven to the plane of the disk by repeated star-disk interactions \citep{SyeCla91,Rau95}. ASC estimate that if the orbital plane of RG is coplanar with the disk, its path inside the disk will be longer by a factor of $\pi r/h$ relative to the perpendicular orbital configuration. Here, $h/r$ is the geometric aspect ratio of the half-thickness of the disk at a given radius and the radius itself. Assuming $h/r\approx 0.1$ for the stellar disk implies that the rate of collisions for an RG can be up to 31 times higher on a coplanar orbit than on a perpendicular orbit. Most RGs should have orbits at intermediate inclinations to the clump disk and may therefore have $t_{\rm coll}$ shorter than that in equation~\ref{eq_tcoll} by a factor less than 31. Along similar lines, RGs brought into co-rotation with the disk will impact clumps at lower relative velocities, diminishing the overall effectiveness of collisions \citep{Rau95}. Therefore, unless most RGs within the central 0.5\,pc in the GC can be brought in counter-rotation with the clump disk, the orientation of the RG orbits cannot be the dominant factor that explains their paucity.

\end{itemize}

In order to allow for tens of RG-clump collisions within the O/WR stellar evolutionary phase, the collision time scale shown in equation~\ref{eq_tcoll} would need to be shorter by a factor of at least $\sim 10^4$. As mentioned previously this can be achieved if the average mass and number of clumps were larger in the past, implying the initial mass of the fragmenting disk that is at least $\sim10^2-10^3$ times larger than that of the presently observed young stellar disk in the GC. For example, based merely on scaling of $t_{\rm coll}$ with $M_c$ and $N_c$, this would require increasing only $M_c$ by 2 orders of magnitude or increasing each $M_c$ and $N_c$ by 1 and 2 orders of magnitude, respectively. Whether the clump number and their mass function can combine to reduce $t_{\rm coll}$ to such a degree is an open question. Given significant uncertainties in the fragmentation and star formation history of the young stellar disk in the GC, this possibility cannot be ruled out.

\subsection{Does mass loss produce less luminous RGs?}

One important implicit assumption made by ASC and adopted here is that the mass loss beyond certain threshold can render RGs in the GC invisible to the current observations. For example, if the entire envelope of an RG star is stripped, the remnant white dwarf like core is expected to be significantly less luminous, especially in the K-band. The stripping of RGs through multiple star-clump collisions is however likely to be gradual, producing a distribution of cores that retain between 0 and 100\% of their envelope. A relevant question for such population is: at which point a stripped RG drops below the sensitivity threshold of existing observational surveys?  

The answer to this question requires calculation of evolutionary tracks of stripped RGs which are beyond the scope of this work. Note however that partial loss of its envelope may not necessarily result in dimming of an RG. For example, \citet{dray2006young} calculate evolutionary tracks of stripped intermediate mass giants ($3-8\,M_{\odot}$) via the tidal forces produced by a SMBH. They show that depending on the exact evolutionary stage in which the stripping happens and the fraction of the envelope remaining, the post-stripping luminosity of the giant star can be either larger or smaller than the pre-stripping luminosity and that it is changing with time. While giants considered by \citet{dray2006young} are more massive than RGs in this work, their results illustrate that the answer to the question of observability of stripped RGs is likely to be nuanced and depend on additional factors besides the mass loss fraction.

One way to sidestep the complexity of gradual stripping of the RG envelopes is by setting a requirement that envelopes of most RGs must be obliterated in a single collision. Our simulations indicate that such conditions are achieved for high clump column densities of $\Sigma_c > 10^{8}\, {\rm g\,cm^{-2}}$ and collision velocities $v_* \gtrsim 900\,{\rm km\,s^{-1}}$.  In the aftermath of such impact the RG star becomes severely distorted and loses a significant amount of its kinetic energy and mass.

\subsection{Simplifying assumptions and their implications}

While we also investigated scenarios in which most of the RG envelope is disrupted in a single impact (mentioned in the previous paragraph), we do not include them in Table~\ref{tab:massloss}, because their simulated mass loss is not reliable. This is because our polytropic models of RG stars capture the density profile of the extended envelope but do not account for the presence of the point-like compact core (see Section~\ref{sub:SettingupRG}). The compact core makes RG more resilient to stripping, relative to the envelope-only model, once $\sim 50\%$ or more of the stellar mass has already been removed. This is because the layers of the star close to the compact core are more strongly gravitationally bound and difficult to remove. The envelope-only stellar model (used in this work) provides a reliable measure of the envelope stripping for scenarios with moderate mass loss, when most of the stellar mass is still enclosed within the stripping radius. 

It is worth noting that when the requirement that tens of RG-clump collisions occur within several Myr is satisfied, the average time between collisions becomes comparable to the thermal time scale of RG stars estimated in equation~\ref{eq_tkh}. This coincidence has interesting implications, particularly in cases when $t_{\rm coll} > t_{\rm KH}$, implying that RG had a sufficient time to radiate away any excess thermal energy deposited by the previous collision before it encounters the next clump. As a consequence, the layers of such thermally relaxed stars are cooler, more tightly bound and harder to strip. Our simulations do not capture thermal relaxation of RGs (as they do not account for radiative processes) and in this case mass loss values reported in this study should be considered as an upper limit. Our computational setup is therefore a more faithful representation of $t_{\rm coll} \lesssim t_{\rm KH}$ scenario.

Our simulations indicate that collisions capable of stripping a non-negligible amount of RG stellar mass also tend to decrease the kinetic energy of a star per impact by a comparable percentage. It follows that any significant amount of stripping of the RG population in the GC must inevitably be mirrored by evolution and systematic decay of their stellar orbits \citep{SyeCla91, Rau95, KarSub01, vilkoviskij2002role}. As a consequence, the density of stars (RG remnants) in the central cluster is expected to steadily increase until the process of migration is counteracted by stellar collisions at cluster radii $\lesssim 10^{-2}\,{\rm pc}$ \citep[equation~16 in][]{ArmZur96}. Note that the SMBH in the Galactic Center is too massive to disrupt the RG cores, even if they are "transported" by the disk close to its event horizon. The tides from the SMBH can however help with stripping of the envelopes of RGs delivered within the central $\sim 10^{-4}\,{\rm pc}$ \citep[equation~1 in][for e.g.]{bogdanovic14}. The RGs subject to grinding by an accretion disk and tidal stripping by the SMBH will therefore reach the center of the cluster as compact remnants.

An additional simplification used in this work is that the gas clumps are modeled as slabs of uniform density, rather then as discrete clouds with some characteristic density profile. Because the clumps are more spatially extended than RG stars by a few orders of magnitude, majority of collisions should happen off-center with respect to the clump core. Such asymmetric collisions can induce rotational velocity change in the star by exerting a torque on the surface of the star. This happens because the density gradient within the clump drives a nonuniform change in the linear momentum across the surface of the star. If most of the change in the linear momentum during the impact (shown in Table~\ref{tab:massloss}) is imparted to one side of RG, then the maximum rotational velocity of the star will be comparable to the escape velocity at the critical (stripping) radius inside the star, as defined by equation~\ref{eq_sigma}. Therefore, 
\begin{equation}
v_{\rm rot} < v_{\rm es} (R_{\rm crit}) \approx 140\,{\rm km\,s^{-1}} 
\left(\frac{R_{\rm crit}}{3 R_\odot}\right)^{-1/2}
\left(\frac{M_*}{M_\odot}\right)^{1/2}
\label{eq_vrot}
\end{equation}
Note that multiple collisions are not necessarily going to torque the star coherently (leading to spin-up or spin-down) and hence the upper limit on the rotational velocity. Spectroscopic studies of G and K giants unaffected by collisions show that the rotational velocity at their surface is about $10\,{\rm km\,s^{-1}}$ \citep{hekker07}. Therefore, enhanced rotational velocity may serve a smoking gun of an RG population affected by collisions.   

Note that clumps are commonly not destroyed in collisions with RGs, even after multiple encounters. The binding energy of a clump with the properties considered in this work amounts to $E_g \approx GM_c^2/R_c \sim10^{49-50}{\rm erg}$, which is orders of magnitude larger than the kinetic energy per impact for most encounters listed in Table~\ref{tab:massloss}. The clumps would however be vulnerable to disruption by the SMBH tides within the radius $r_t \approx R_c \,(M_\bullet/M_c)^{1/3} \sim 10^{15}\,{\rm cm}$ from the center of the Galaxy. Therefore, star-disk collisions cannot be the mechanism responsible for the scarcity of RGs within the central $\sim 10^{-3}\,{\rm pc}$. It is within this region however that the SMBH tides become strong enough to strip the RG envelopes, as discussed earlier in this section.

The simulations presented here are purely hydrodynamic and do not take into account ambient magnetic fields that may be present in the fragmenting, proto-stellar disk. The role of magnetic fields has previously been investigated in scenarios that involve overdense gas clouds (rather then stars) impacted by magnetized shock fronts in the interstellar medium \citep{1994ApJ...433..757M, 2005ApJ...619..327F, 2007ApJ...670..221D, 2008ApJ...680..336S}. These studies find that magnetic field lines draped over the cloud tend to suppress the growth of destructive hydrodynamic instabilities and protect it from fragmentation. If instead of a gas cloud one considers a supersonic impact of RG with a magnetized clump, these findings suggest that presence of magnetic fields could help to shield the RG star from stripping.


\section{Conclusions}
\label{sec:conc}

We investigate the hypothesis that collisions of stars with a fragmenting accretion disk are responsible for the observed dearth of RG and HB type stars within the central 0.5\,pc of the Galactic Center. In this context, we model evolved stars representative of the missing population as they collide with dense clumps of gas and establish physical conditions under which stars can lose a significant fraction of their initial mass. Our main findings are as follows.
\begin{itemize}
\item Substantial mass loss ($\gtrsim 10\%$) is possible for stars that experience multiple collisions with clumps of column densities $\gtrsim10^{8}\, {\rm g\,cm^{-2}}$. Such high column densities are characteristic of the accretion disks in the late stage of fragmentation and on the verge of forming stars. 

\item Repeated star-clump impacts are found to be particularly efficient at stripping the RG envelopes, as predicted by \citet{ArmZur96}. This is because repeated impacts and incremental mass loss drive expansion of the star between collisions and increase its cross section in subsequent encounters. Comparing simulations where the star is tunneling through a continuous medium to those with repeated impacts, we find that the latter show systematically larger changes in the momentum and kinetic energy of the star, as well as the mass loss per impact. 

\item We compare the mass loss measured in simulations with analytic expectations and find them to be in reasonable agreement (within a factor of 10), whenever simulated mass loss is averaged over more than a few impacts. For most runs, the analytic expectations underestimate the values of mass loss measured from simulations. We also use simulations to calibrate the mass loss factor which can be used to obtain an empirical estimate of the mass loss for an arbitrary number of impacts, as long as the star is losing mass in a steady, non-runaway fashion.

\item To investigate the effect of stellar structure on mass loss we model RG stars as $\Gamma=5/3$ and $\Gamma=4/3$ polytropes. We find that the simulated mass loss is comparable for the two models to within 30\% with $\Gamma=4/3$ polytropes resulting in a smaller mass loss. While RGs modeled as the $\Gamma=4/3$ polytropes are vulnerable to stripping to a much smaller radius within the star, their tenuous envelopes contain only a small fraction of the total mass and the combination of these two effects results in a smaller overall mass loss. We therefore consider the value of the mass loss measured from the $\Gamma=5/3$ polytropes as an upper bound for both models.

\item Collisions strip the RG envelope on account of the kinetic energy of the star, causing it to drop by the percentage comparable and slightly larger than the mass loss percentage. Therefore, any significant amount of stripping of the RG population in the GC must be mirrored by a systematic decay of their stellar orbits. Along similar lines, collisions with clumps can induce rotational velocity change in the star by torquing the surface of the star. Isolated G and K giants are relatively slow rotators and therefore, enhanced rotational velocity may serve a smoking gun of an RG population affected by collisions.

\item In order to allow for multiple RG star-clump collisions within the several Myr long phase during which clumps are abundant, the total mass of the fragmenting disk must be $\sim2-3$ orders of magnitude higher than that of the WR/O stars which now form the stellar disk in the GC. While we cannot determine the plausibility of that physical scenario based on this work, such possibility cannot be ruled out given significant uncertainties in the fragmentation and star formation history of the young stellar disk in the GC.
\end{itemize}


\acknowledgments 

The authors thank an anonymous referee for a thoughtful and useful report. T.F.K. gratefully acknowledges support from the Barry Goldwater Scholarship and Presidents Undergraduate Research Awards (Georgia Institute of Technology). The authors thank Roseanne Cheng for providing her version of the VH-1 code for this study and Tal Alexander, Pau Amaro-Seoane, Matt Benacquista, Melvyn Davis, Cole Miller, and Enrico Ramirez-Ruiz for their insightful comments.  T.B. acknowledges the support from the Alfred P. Sloan Foundation under Grant No. BR2013-016 and the National Science Foundation under grant No. NSF AST-1333360. Numerical simulations presented in this paper were performed using the high-performance computing cluster PACE, administered by the Office of Information and Technology at the Georgia Institute of Technology.


\bibliographystyle{apj}
\bibliography{apj-jour,myrefs}


\appendix
\label{appendix}

\section{Numerical convergence and stability of the star}
\label{sec:resolutionstudy}

In this section we describe the criteria for numerical convergence and uncertainties associated with the finite numerical resolution employed in our simulations. As mentioned in Section~\ref{sub:SettingupRG} we model RGs as the $\Gamma = 5/3$ and $\Gamma = 4/3$ polytropes in order to investigate the effect of the stellar structure on mass loss. We choose numerical resolution $128^3$ for the $\Gamma = 5/3$ model and prior to simulating RG-clump collisions verify that this resolution is sufficient to capture the structure of the star over the time scales relevant to this study.

\begin{figure*}[!b]
\begin{center}
\includegraphics[scale=0.7]{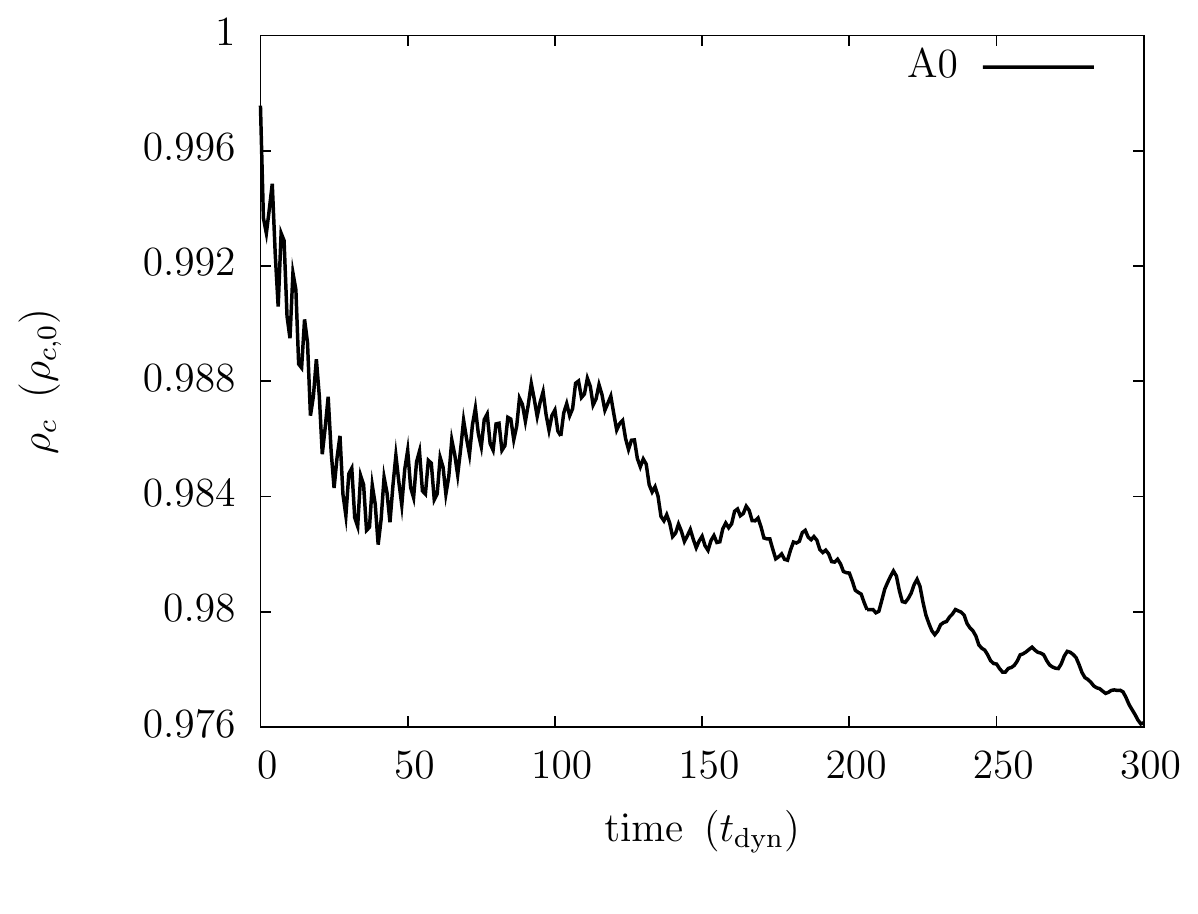}
\includegraphics[scale=0.7]{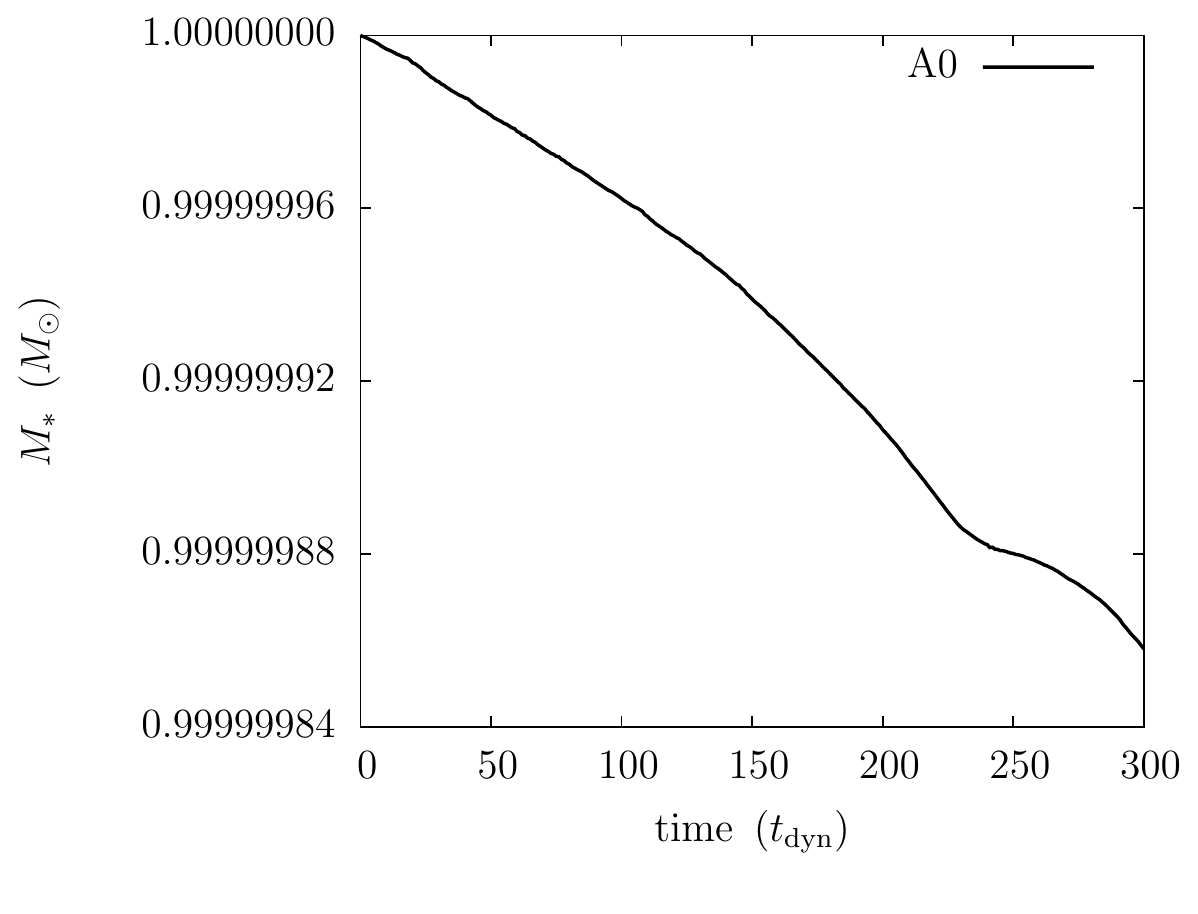}
\caption{Central density (left) and mass (right) of RG as a function of time for run A0.  We require that the evolution of the star in the baseline simulation is slow and that the central density remains above $0.97\, \rho_{c,0}$. This stability criterion is satisfied during the initial $\sim 300\, t_{\rm dyn}$ at numerical resolution of $128^3$.}
\label{fig:ResolutionStudy_A0}
\end{center}
\end{figure*}

Figure~\ref{fig:ResolutionStudy_A0} shows the evolution of the central density and mass of the star in run A0 (shown in Table~\ref{tab:A0}). In this run the star is tunneling through a smooth gaseous accretion disk, which properties correspond to a marginally stable disk described by \citet{levin07} at a radius $\sim0.02$\,pc. Because of its very low surface density, compared to the star and the clumps in the fragmenting disk, the perturbation in the central density of the star and mass loss are negligible. Most of the evolution seen in Figure~\ref{fig:ResolutionStudy_A0} can actually be attributed to the expansion of the star due to numerical diffusion, caused by the finite numerical resolution. We therefore use the A0 run as a test of numerical stability of the star placed in a low density background flow. We require that the evolution of the star in the baseline simulation is slow and that central density remains above $0.97 \rho_{c,0}$. This criterion is satisfied during the initial $300\, t_{\rm dyn}$ at numerical resolution of $128^3$ and we apply it to the remainder of simulations in this work (in all simulations we do not consider the data beyond $300\, t_{\rm dyn}$). We use the mass loss of $1.4\times 10^{-7}\,M_\odot$ recorded during $300\, t_{\rm dyn}$ in the A0 simulation as an estimate of the absolute error in the mass loss measurements for all simulations (i.e., the smallest mass loss that can be reliably measured) due to the approximate stability of the star.

\begin{deluxetable*}{cccccccccccc}[!t]
\tabletypesize{\footnotesize} 
\tablecolumns{12} 
\tablewidth{0pt} 
\tablecaption{A0 run parameters. $\Gamma$ -- polytropic index. $N_{\rm res}$ -- numerical resolution. $v_*$ -- velocity of the star. $\Sigma$ -- disk column density. $h$ -- disk half-height. $\mathcal{M}$ -- Mach number. $t_{\rm dc}$ -- disk crossing time.  $t_{\rm sim}$ -- simulation length. $N_{\rm coll}$ -- number of collisions. $\Delta M_a$ -- estimated mass loss over 300\,$t_{\rm dyn}$. $\Delta M$ -- mass loss measured in simulations over 300\,$t_{\rm dyn}$. \label{tab:A0}} 
\tablehead{ 
Run  & $\Gamma$ & $N_{\rm res}$ & $v_{*}$ & $\Sigma$ & $h$ &  $\mathcal{M}$ & $t_{\rm dc}$ & $t_{\text{sim}}$ & $N_{\rm coll}$  & $\Delta M_a$ & $\Delta M$\\
& & & (km\,s$^{-1})$  & (g\,cm$^{-2}$)  & (cm) & & ($t_{\text{dyn}}$)  &  ($t_{\text{dyn}}$) & & ($M_\odot$) & ($M_\odot$)}
\startdata 
A0  & 5/3 & 128 & 300 & $2472$ & $6.8 \times 10^{15}$  & 284 & 4502 & 300 & $<$1 & $\approx 0$ & $1.4 \times 10^{-7}$ \\  
\enddata  
\end{deluxetable*}

\begin{figure*}[!h]
\begin{center}
\includegraphics[scale=0.7]{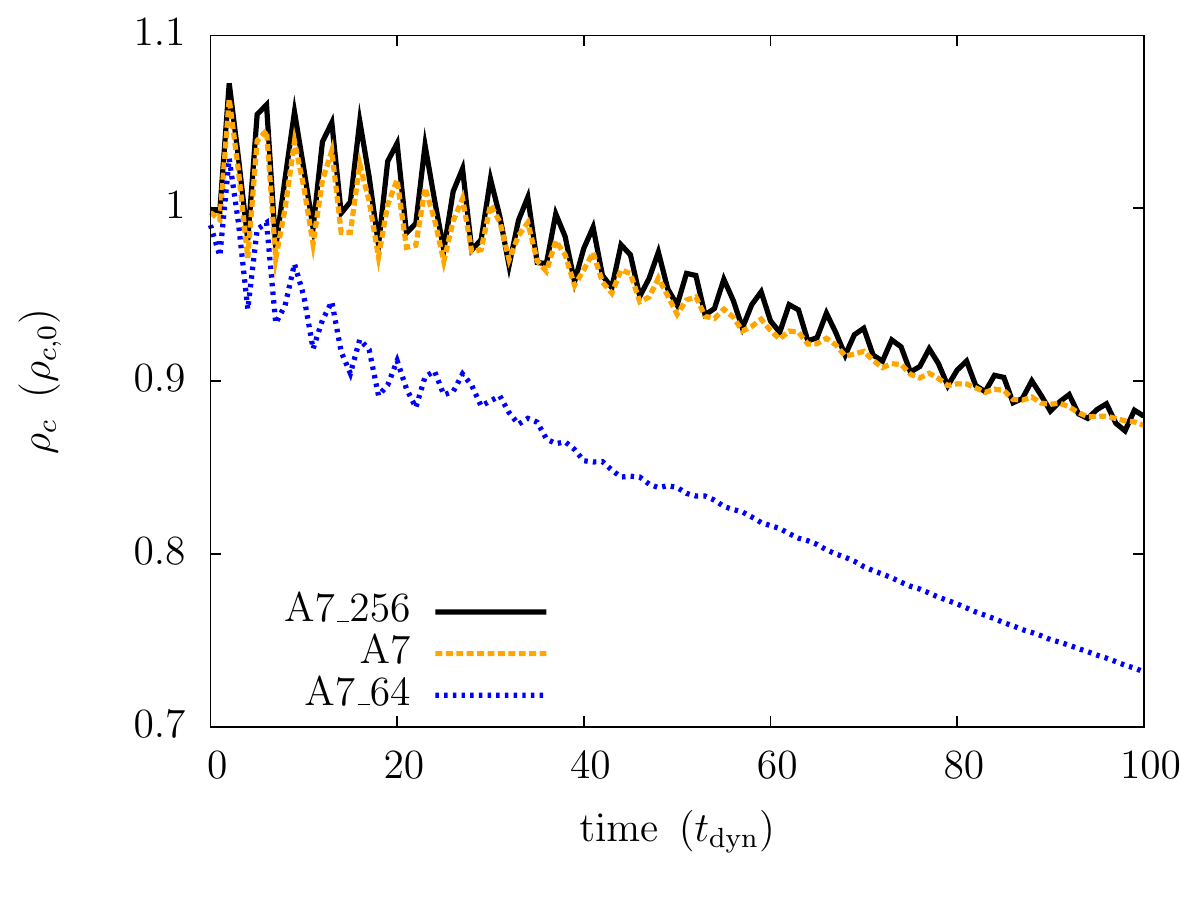}
\includegraphics[scale=0.7]{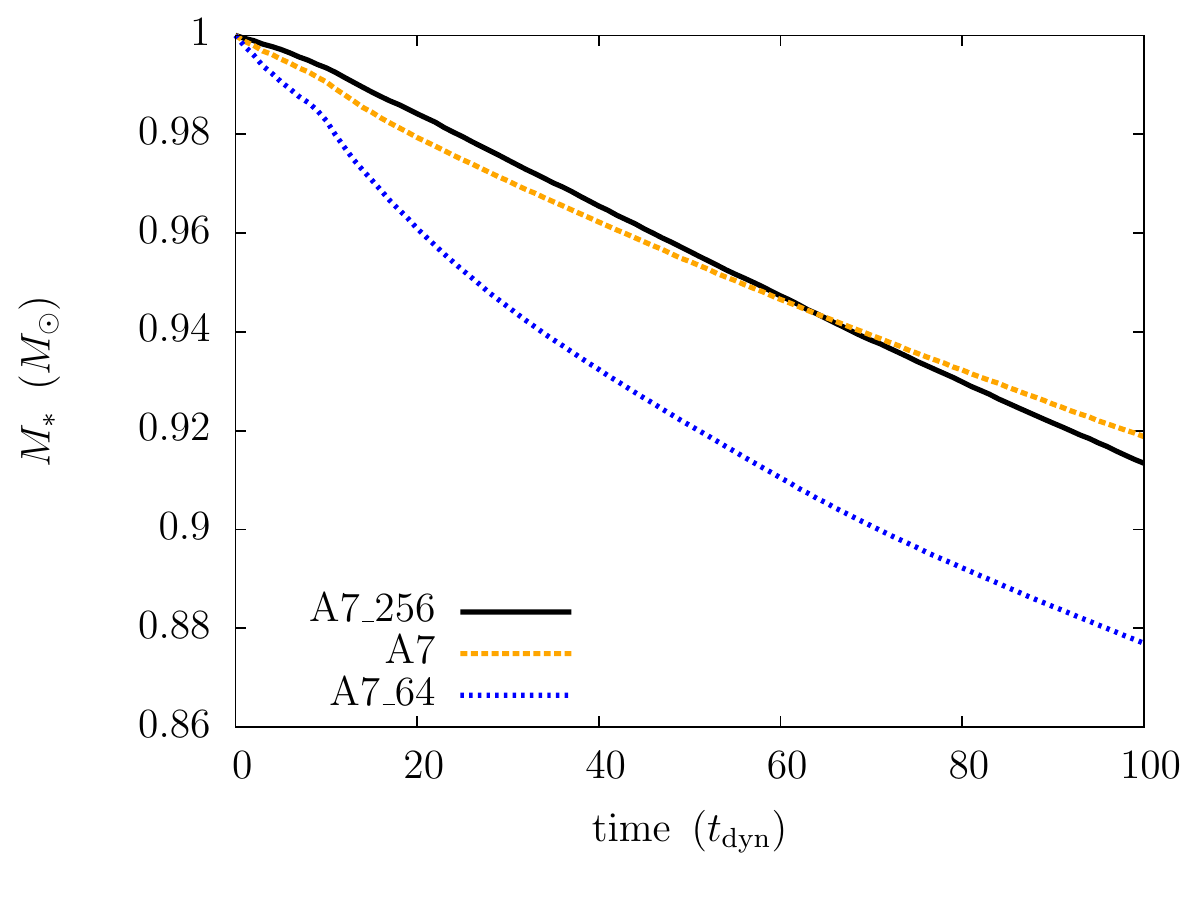}
\caption{Evolution of the central density (left) and mass of the star (right) as a function of numerical resolution for runs A7\_64, A7, and A7\_256. Simulations with resolutions $128^3$ and $256^3$ converge to comparable values and show relative difference in the mass loss corresponding to 6\% after $100\, t_{\rm dyn}$. Different line styles correspond to resolutions of $64^3$ (blue, dotted), $128^3$ (yellow, dashed), and $256^3$ (black, solid). \label{fig:ResolutionStudy_A7}}
\end{center}
\end{figure*}

We also examine the central density and mass loss as a function of numerical resolution as a test of numerical convergence in our runs. Figure~\ref{fig:ResolutionStudy_A7} shows the numerical convergence test for run A7 ($\Gamma = 5/3$ polytrope). The numerical resolution of $64^3$ is insufficient to properly resolve the pressure and density gradients within the star and as a consequence the central density drops precipitously, the star expands and experiences increased mass loss. The simulations with resolutions $128^3$ and $256^3$ converge to comparable values and have two noticeable differences. Firstly, oscillations in the central density of the star decay more quickly in the lower resolution run A7, which is expected to be more diffusive. Secondly, the difference in the mass loss between the runs A7 and A7\_256 over $100\, t_{\rm dyn}$ is about $5 \times 10^{-3}$. We use this measurement to establish a relative error in the mass loss measurements caused by the finite numerical resolution as $\Delta M = (8.1 \pm 0.5)\times 10^{-2}\,M_\odot$, which corresponds to about 6\%.

\begin{figure*}[!t]
\begin{center}
\includegraphics[scale=0.7]{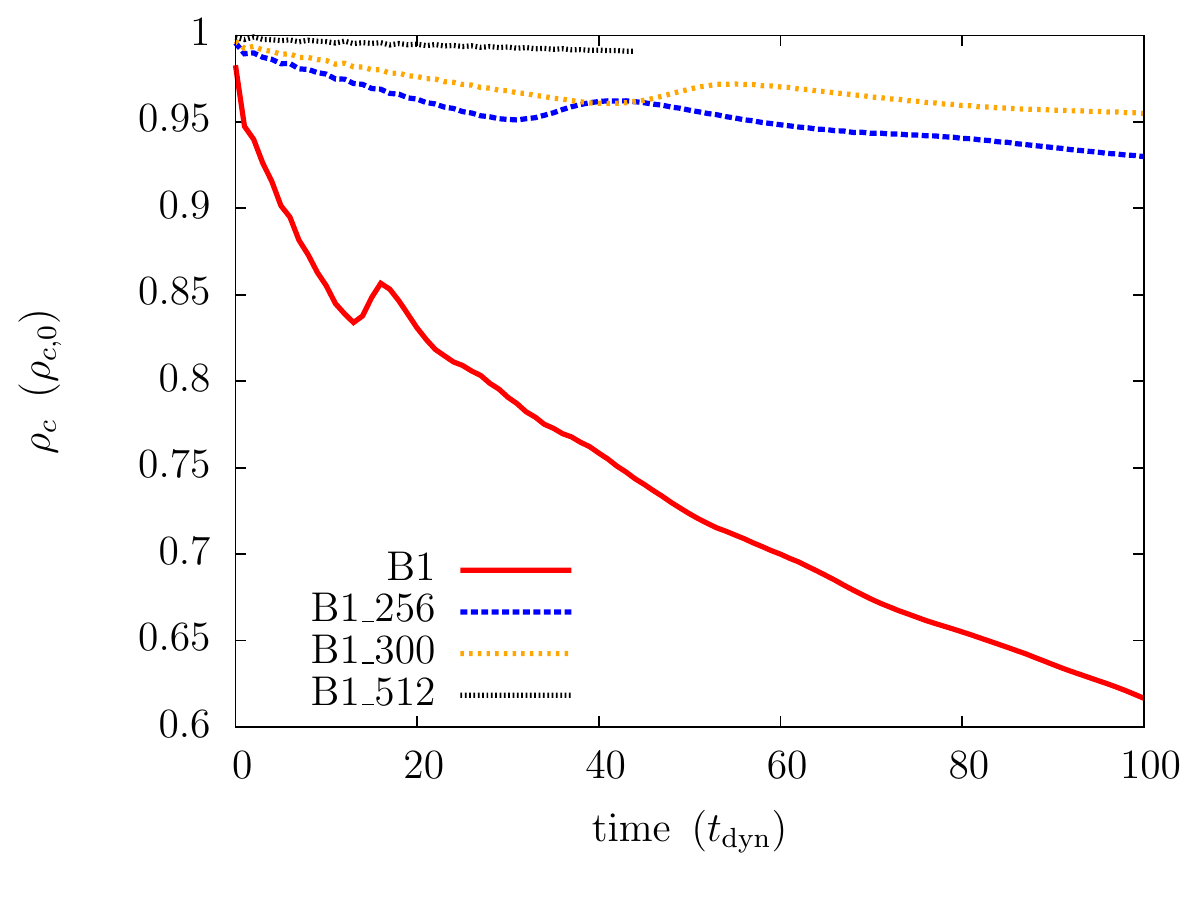}
\includegraphics[scale=0.7]{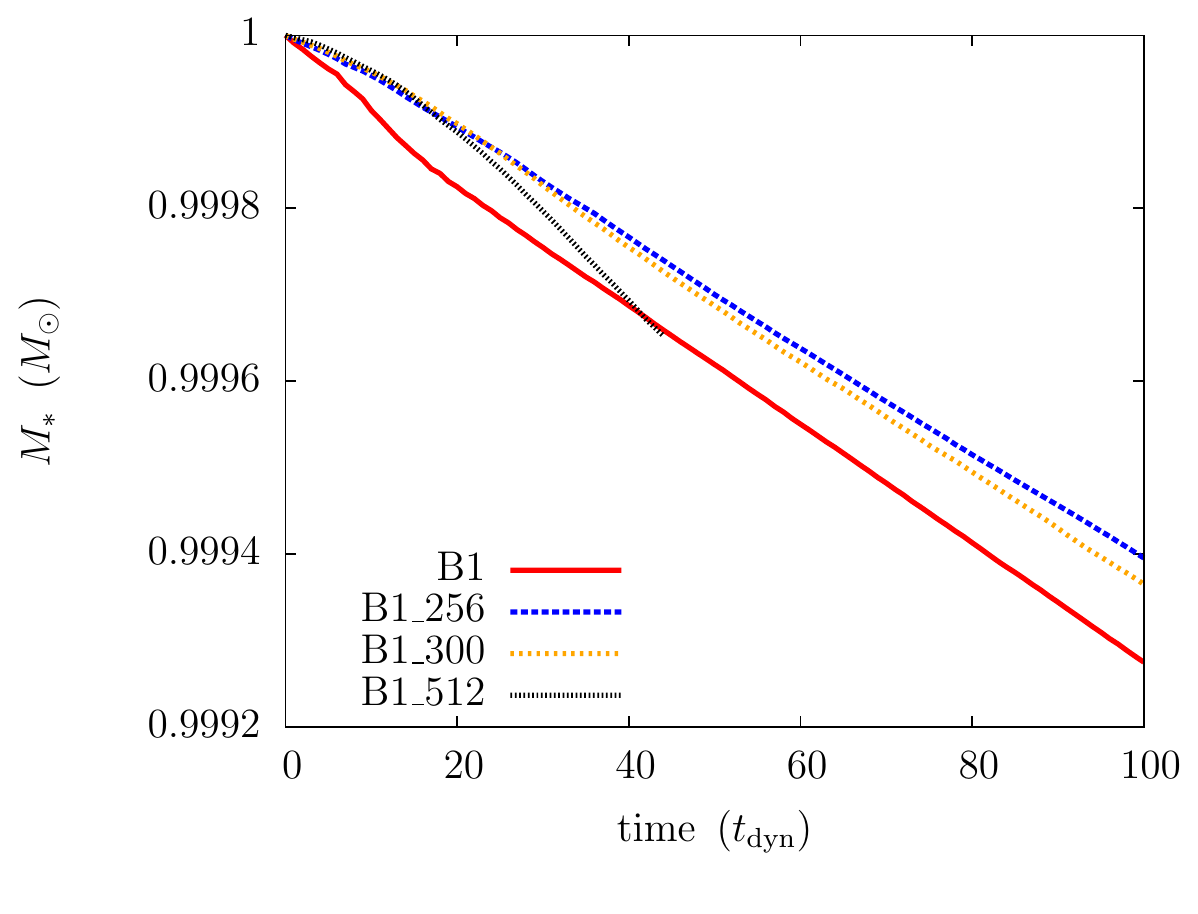}
\caption{Evolution of the central density (left) and mass of the star (right) as a function of numerical resolution for runs B1, B1\_256, B1\_300, and B1\_512. Different line styles correspond to resolutions of $128^3$ (red, solid), $256^3$ (blue, dashed), $300^3$ (yellow, dotted), and $512^3$ (black, dotted).  Simulation with resolution $512^3$ is computationally expensive and we only run it for about $40\,t_{\rm dyn}$. Simulations with resolutions $300^3$ and $512^3$ show a a few percent difference in the central density and 0.01\% difference in the mass loss. \label{fig:ResolutionStudy_B1}}
\end{center}
\end{figure*}

Figure~\ref{fig:ResolutionStudy_B1} shows the numerical convergence test for run B1 ($\Gamma = 4/3$ polytrope). Because they have much steeper central pressure and density gradients stars modeled as $\Gamma = 4/3$ polytropes require higher numerical resolution relative to the $\Gamma = 5/3$ polytropes. We show the central density and mass of RG in run B1 for numerical resolutions of $128^3$, $256^3$, $300^3$, and $512^3$. In this case, numerical resolution of $128^3$ is insufficient to capture the structure of the star, which quickly dissolves. The evolution of the star appears similar for resolutions $256^3$ and $300^3$ and we choose the latter as the standard resolution in our simulations of $\Gamma = 4/3$ polytropes. Note that increased numerical resolution makes these runs $(300/128)^4 \approx 30$ times more computationally expensive than those at $128^3$ and we carry out fewer of these runs. We also carried out a shorter simulation at resolution of $512^3$ and note that B1\_512 exhibits less diffusive behavior when it comes to the central density and differences at the level of $\sim 10^{-4}\,M_\odot$ in the mass loss. Note that the compact core in the $\Gamma = 4/3$ polytropic model remains largely unperturbed in this scenario and in contrast to the $\Gamma = 5/3$ models, there is no "ringing" in the central density. 

\end{document}